\documentclass[journal, 12pt, draftclsnofoot, onecolumn]{IEEEtran}
\IEEEoverridecommandlockouts

\usepackage{amsfonts,amssymb}
\usepackage{ifpdf}
\usepackage{cite}
\usepackage{stfloats}
\usepackage{bm}
\usepackage{color,xcolor}
\usepackage{subfigure}
\ifCLASSINFOpdf
   \usepackage[pdftex]{graphicx}
   \graphicspath{{../pdf/}{../jpeg/}}
   \DeclareGraphicsExtensions{.pdf,.jpeg,.png}
\else
   \usepackage[dvips]{graphicx}

   \graphicspath{{../eps/}}

   \DeclareGraphicsExtensions{.eps}
\fi

\usepackage[cmex10]{amsmath}

\usepackage{algorithm}
\usepackage{algorithmic}
\ifCLASSOPTIONcompsoc
\else
  \usepackage[caption=false,font=footnotesize]{subfig}
\fi

\usepackage{makecell}
\begin{document}
\title{UAV-Assisted Intelligent Reflecting Surface Symbiotic Radio System}
	
\author{Meng~Hua,
Luxi~Yang,~\IEEEmembership{Senior Member,~IEEE,}
Qingqing~Wu,
Cunhua~Pan,
Chunguo~Li,~\IEEEmembership{Senior Member,~IEEE,}
and~A. Lee Swindlehurst,~\IEEEmembership{Fellow,~IEEE}
%
%%\thanks{Copyright (c) 2015 IEEE. Personal use of this material is permitted. However, permission to use this material for any other purposes must be obtained from the IEEE by sending a request to pubs-permissions@ieee.org.}
%\thanks{Manuscript received April   27, 2019; revised July    29, and accepted November  7, 2019. This work was supported by National Natural Science Foundation of China under Grant  61971128,  Grant 61372101, and Grant 61671144, Scientific Research Foundation of Graduate School of Southeast University  under Grand  YBPY1859 and China Scholarship Council (CSC) Scholarship, National High Technology Project of China  under 2015AA01A703,  Cyrus Tang Foundation Endowed Young Scholar Program under SEU-CyrusTang-201801.   The associate editor coordinating the review of this paper and approving it for publication was Kamel Tourki. (\emph{Corresponding author: Luxi Yang}.)}
\thanks{M. Hua is with the School of Information Science and Engineering, Southeast University, Nanjing 210096, China, and  also  with the State Key Laboratory of Internet of Things for Smart City and Department of Electrical and Computer Engineering, University of Macau, Macao 999078 China  (e-mail: mhua@seu.edu.cn).}
\thanks{   L. Yang and C. Li are with the School of Information Science and Engineering, Southeast University, Nanjing 210096, China (e-mail: \{  lxyang, chunguoli\}@seu.edu.cn).}
\thanks{Q. Wu is with the State Key Laboratory of Internet of Things for Smart City and Department of Electrical and Computer Engineering, University of Macau, Macao 999078 China (email: qingqingwu@um.edu.mo). }
\thanks{C. Pan is with the School of Electronic Engineering	and Computer Science, Queen Mary University of London, London E1 4NS, U.K. (e-mail: c.pan@qmul.ac.uk).}
\thanks{A. L. Swindlehurst is with the Center for Pervasive Communications and Computing, University of California at Irvine, Irvine, CA 92697 USA (e-mail: swindle@uci.edu).}
%\thanks{Part of this work has been accepted by IEEE Global Communications Conference 2019 \cite{Hua2019throughput}. }.
}
\maketitle
\vspace{-2em}
\begin{abstract}
This paper investigates a symbiotic unmanned aerial vehicle (UAV)-assisted intelligent reflecting surface (IRS) radio system, where the UAV is leveraged to help the IRS reflect its own signals to the base station, and meanwhile enhance the UAV transmission by passive beamforming  at the IRS.
First, we consider the weighted sum bit error rate (BER)  minimization  problem  among all IRSs  by jointly  optimizing the UAV trajectory, IRS phase shift matrix, and IRS scheduling, subject to  the minimum primary rate requirements.  To tackle this complicated  problem, a relaxation-based algorithm is proposed.  We prove that the converged relaxation scheduling variables are binary, which means that no reconstruct strategy is needed, and thus the UAV rate constraints  are automatically  satisfied. Second, we consider the fairness BER optimization  problem.  We find that the  relaxation-based  method cannot  solve  this fairness BER  problem since the minimum primary rate requirements may not be satisfied by the binary reconstruction operation. To address this issue, we first transform the binary constraints  into  a  series of equivalent equality constraints. Then, a  penalty-based algorithm is proposed to obtain a  suboptimal solution. Numerical results are provided to evaluate the performance of the proposed designs under different setups, as compared with benchmarks.
\end{abstract}


\section{Introduction}
With the ever-growing sales of mobile devices and Internet of Things devices, current network architectures are becoming overwhelmed by growing data traffic demands \cite{wu2017anoverview}. Although numerous  technologies such as   millimeter wave (mmWave) communications, ultra-dense networks, and  massive multiple-input multiple-output (MIMO) \cite{swindlehurst2014millimeter, kamel2016ultra,lu2014overview} have been proposed to address this problem, they are   usually realized with very large energy consumption and high hardware cost due to the large number of RF chains required at the terminals. Recently, a new technology has come to the attention of the wireless research community, namely intelligent reflecting surface (IRS), due to its potential ability to reconfigure the radio propagation environment in a favorable way for  transceiver optimization. An IRS is comprised of a manmade
surface of electromagnetic  material consisting of a large number of square metallic patches, each of which can be digitally controlled to induce  different reflection  amplitude, phase, and  polarization responses on the incident signals \cite{WU2020towards}, \cite{cui2017information}.  Since an IRS typically has numerous patch units (such as PIN-diodes), it can provide a significant passive beamforming gain without the need for RF chains, thus yielding a cost- and energy-efficient solution.  For example,  experiments conducted recently in \cite{tang2019wireless} showed that for a large IRS consisting of $1720$ reflecting elements, the total power consumption  is only  $0.280 \rm W$.  In addition, each IRS reflecting element adjusted  by the smart controller is able to induce an independent phase shift on the incident signal to change the signal propagation such that the desired and interfering signals can be added constructively or destructively to assist the communication system.
Therefore, IRS is a promising solution for improving the spectral and energy efficiency of wireless networks, and paving the way to the green networks of the future.

The new research paradigm of IRS-aided wireless communication has been extensively studied, e.g., see  \cite{Liaskos2018anew,marco2020smart,wu2020intelligentarxiv}. The authors of \cite{Liaskos2018anew}  proposed a radically software control approach, to enable   programmable control over the behavior of IRS-related wireless environments.
 How the availability of IRS will allow wireless network operators to redesign common and well-known  network communication paradigms was discussed in \cite{marco2020smart}.
The authors of  \cite{wu2020intelligentarxiv} provided an overview of the promising IRS technology for achieving smart and reconfigurable environments in future wireless networks, and  elaborated the reflection and channel models, hardware architecture  as well as various applications.

 Recently, there have been many contributions devoting efforts to integrating IRS into the current cellular networks.  
Joint active and passive beamforming design was  investigated to either maximize the system throughput or minimize the base station (BS) transmit power in \cite{wu2019intelligent,huang2020ReconfigurableIntelligent,pan2019multicell,huang2019Reconfigurable,zhang2020intelligent}. In particular, the authors in  \cite{wu2019intelligent} studied the BS transmit power minimization problem by jointly optimizing the BS beamforming matrix and IRS phase shift matrix while satisfying the users' minimum signal-to-interference-plus-noise ratio (SINR) requirement, and  the results showed  that for a single-user IRS-aided system, the received signal-to-noise ratio (SNR) increases quadratically with the number of reflecting elements.  In addition, the  applications of IRS are also appealing for numerous different  system setups such as  spectrum sharing \cite{guan2020joint}, physical layer security \cite{yang2020deep,zhou2020framwork,lu2020roubust},  orthogonal multiple access \cite{yang2019intelligent}, \cite{yang2020irs}, and simultaneous wireless information and power transfer \cite{wu2019weighted,pan2019intelligent,li2020joint}.

Unlike the above studies in  which the IRS is used purely to assist the transmissions of the existing system, a new IRS functionality referred to as symbiotic radio transmission has  been  proposed (also known  as passive beamforming and information transfer transmission), where the information bits are carried by the on/off states of the IRS, while passive beamforming is achieved by adjusting the phase shift of each  reflecting element \cite{yan2019passive,zhang2020large,hu2020Reconfigurable}. Specifically, a sensor is integrated into the IRS system,  which for example collects environmental information such as  temperature, humidity, illuminating light, etc., and sends it to a smart controller at the IRS   via adjusting  the  on/off state of the  IRS. Then, the  controller  transmits  the collected  information to the BS by adjusting the on/off state of the IRS.  This concept is similar to the spatial modulation transmission technique, where the active transmit antenna number is regarded as a  source of information to improve  the spectral efficiency \cite{Mesleh2008Spatial}. 
%In fact, this idea has been tested experimentally in \cite{ma2019smart}, where a motion-sensitive smart metasurface was integrated with a three-axis gyroscope on an aircraft to sense its direction of motion, allowing a smart controller to adaptively adjust the IRS phase shifts to maintain a beam pointed at a desired receiver.  

\begin{figure}[!t]
\centerline{\includegraphics[width=3in]{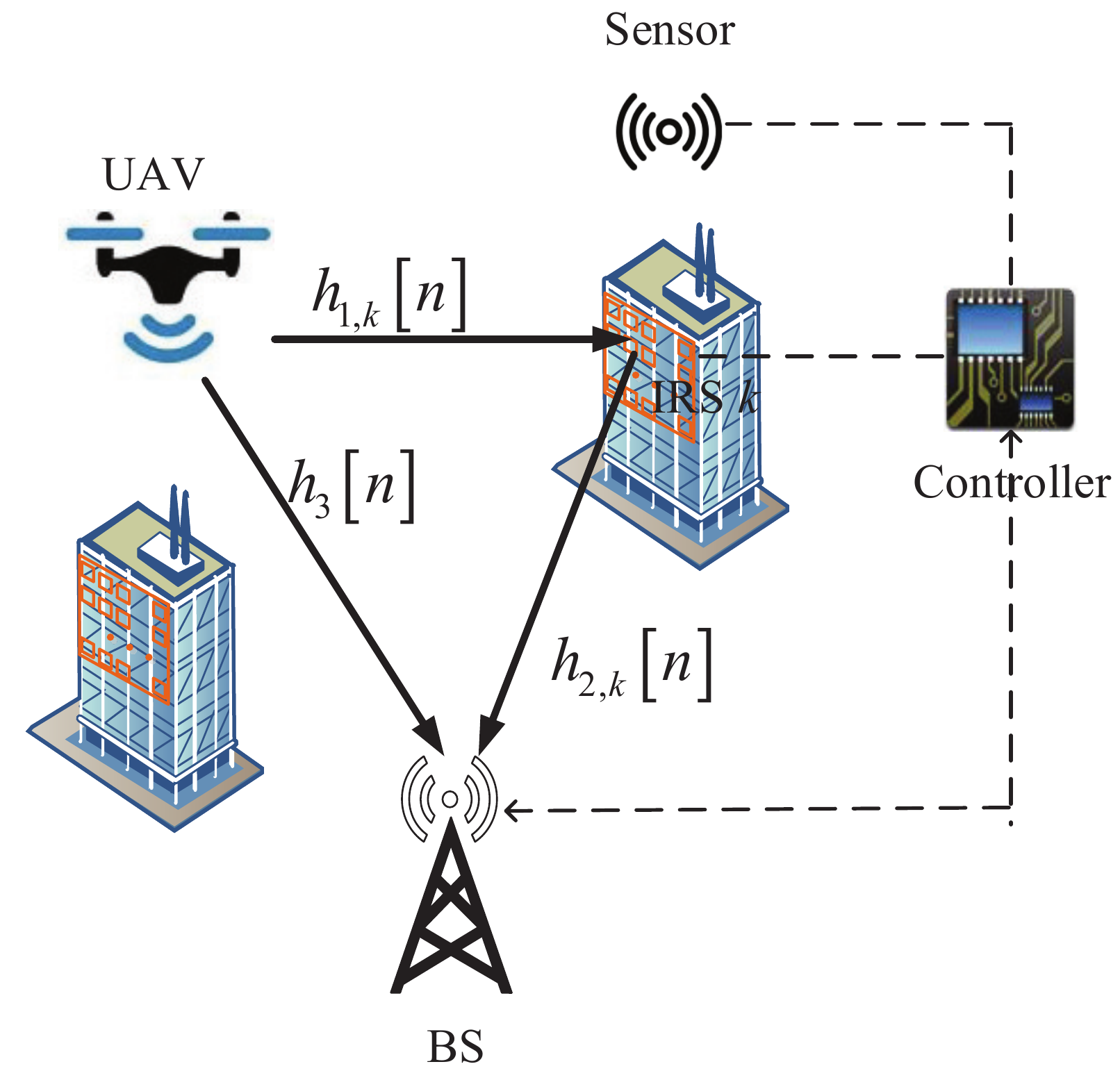}}
\caption{UAV assisted IRS Symbiotic Radio  System.} \label{fig1}
\end{figure}

In this paper, we study an unmanned aerial vehicle (UAV)-assisted IRS symbiotic radio system, where the UAV is leveraged  to assist the IRS data  transmission. Specifically, we consider an urban environment, where there are multiple IRSs available to sense environmental information.  As shown in Fig.~\ref{fig1}, the IRS  sends its own data to the BS by controlling its on/off state, and the receiver side (BS) uses the difference in channel response caused by the on/off state to decode the IRS information. The IRS  also simultaneously tunes each reflecting element to align the phase of the signal passing through the UAV-IRS-BS link with that of the UAV-BS link to achieve coherent signal combining at the BS, thereby enhancing the UAV communication performance. In addition, the UAV’s flexible mobility can be exploited to create favorable channel conditions for the UAV-BS and UAV-IRS links. Our goal in this paper is to minimize the bit error rate (BER) of IRS  by jointly optimizing the UAV trajectory, IRS phase shift matrix, and IRS scheduling, subject to a minimum  data rate requirement for the UAV.  We study two optimization objectives, one based on fairness for the IRS and the other on the weighted sum BER of the IRS. Then, we develop two novel  algorithms to solve them. The main contributions of this paper are summarized as follows:
\begin{itemize}
		\item We first consider the IRS weighted sum BER  optimization problem, which is  a  mixed-integer and non-convex problem.  To  develop a low complexity algorithm, we propose a relaxation-based method. Specifically, we first relax the binary scheduling variables to continuous variables, and then we develop an alternating optimization (AO) algorithm to solve the relaxed  non-convex optimization problem. We prove that the obtained scheduling results  are the same as the binary results from the AO  method, which means that no reconstruct strategy is needed, and thus  the primary rate requirements of UAV are always satisfied. 	
		 Numerical results show  the proposed relaxation-based method converges within only a few iterations.
	
	\item We then take into account the fairness among multiple IRSs, and formulate a fairness BER optimization problem. The resultant problem is also a mixed integer and non-convex problem, which is in general  difficult to solve optimally. We show that the commonly used relaxation-based method cannot be applied to this problem since the UAV's rate  requirement constraints may not be satisfied by the binary reconstruction operation for scheduling. To address this issue, a novel penalty-based algorithm is proposed. We first transform the binary constraints into a series of equivalent equality constraints, and then propose a two-layer algorithm to solve the problem. Numerical results show the effectiveness of this penalty-based algorithm.

	\item We conduct simulation results for the two proposed scenarios to illustrate their performance. For the first  scenario, we study the impact of weighting factors on the system performance, and find that the optimized UAV trajectory places it closest to IRS with a high weighting factor.	For the second scenario,  the results show that the average fairness utility value   is highly related to the IRS phase shift matrix. In addition, for both scenarios, the system performance is  significantly improved by  the optimized UAV trajectory as well as the finely tunable IRS phase shift.
	 
\end{itemize}
The rest of this paper is organized as follows: Section II introduces the system model and problem formulation. In Sections III and IV, we study the  weighted sum BER and fairness BER optimization problems, respectively. Numerical results are provided in Section V, and  the paper is concluded in Section VI.

\emph{Notations}: Boldface lower-case variables denote vectors. The notation ${\left\| {\bf x} \right\|}$ represents the  Euclidean norm of $\bf x$, the circularly symmetric complex Gaussian variable $x$ with mean $ \mu$ and variance  $ \sigma^2$ is denoted by ${x} \sim {\cal CN}\left( {{{\mu }},{{\sigma^2 }}} \right)$, statistical expectation is defined as ${\mathbb E}\left\{  \cdot  \right\}$, statistical variance  is defined as ${\mathbb V {\rm ar}}\left\{  \cdot  \right\}$, and ${\cal O}\left(  \cdot  \right)$ denotes the big-O computational complexity notation.
\begin{figure}[!t]
\centerline{\includegraphics[width=3in]{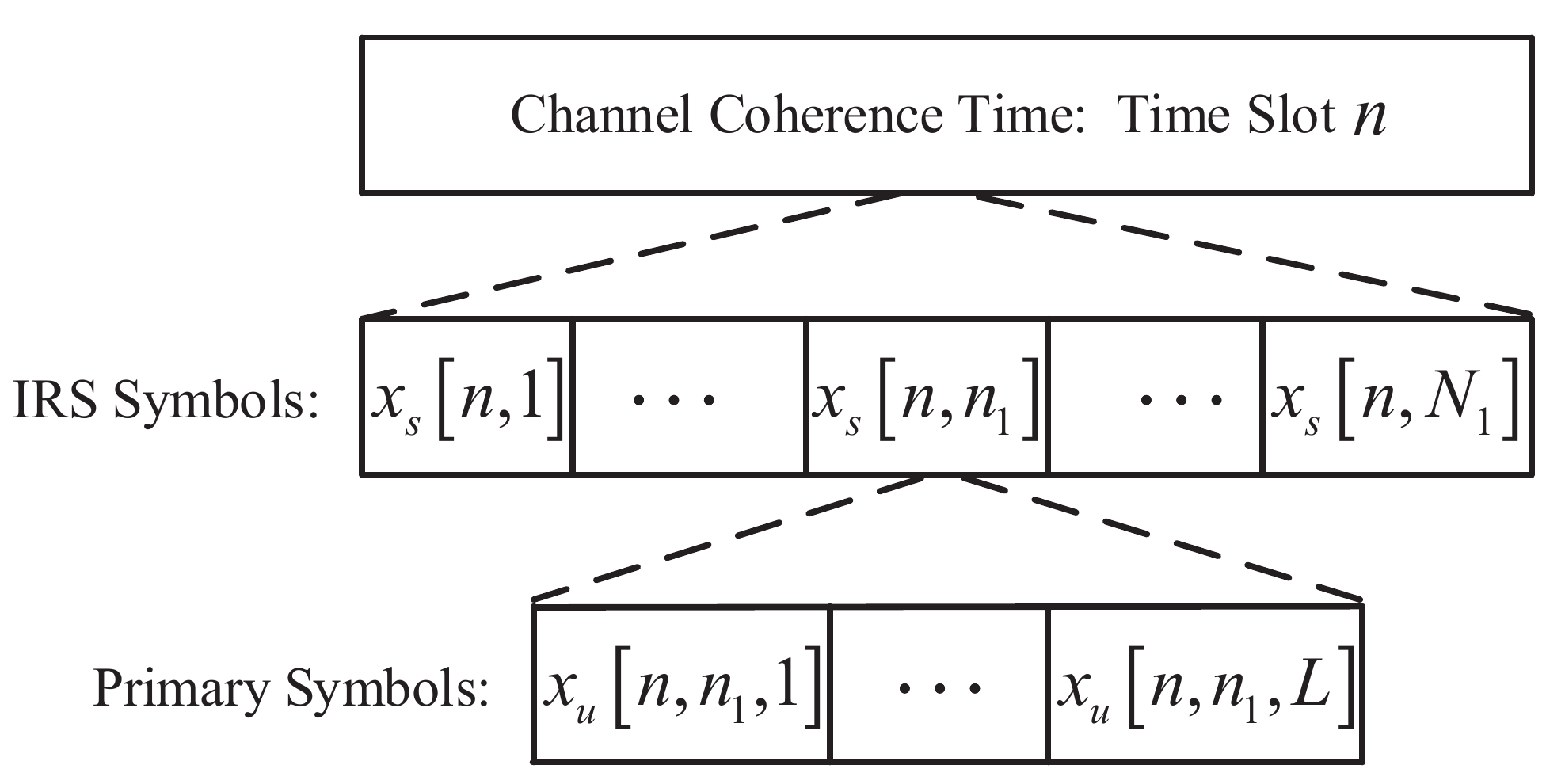}}
\caption{Transmission frame for the IRS and primary transmission.} \label{fig2}
\end{figure}
\section{System Model and Problem Formulation}
\subsection{System Model}
Consider a UAV-assisted IRS symbiotic radio  system consisting of a single-antenna  BS, a single-antenna UAV, and $K$ IRSs as shown in Fig.~\ref{fig1}, where  the UAV acts to help the IRS transmit its own data to the BS\footnote{For ease of exposition, we term the UAV and BS as  a primary network to assist the IRS's information transmission.}.
The BS and IRS are in fixed locations, and the UAV can freely adjust its heading as it moves. The horizontal coordinates of the BS and the $k$th IRS are respectively denoted as ${{\bf{q}}_b} = {\left[ {q_b^x,q_b^y} \right]^T}$ and ${{\bf{q}}_{s,k}} = {\left[ {q_{s,k}^x,q_{s,k}^y} \right]^T}$. In addition, the altitude of the BS is denoted as $H_b$, and the altitudes of all the IRSs are the same denoted as $H_s$\footnote{Note that adopting different altitudes of IRSs do not affect the algorithm design in this paper.}.
We assume that the UAV flies in a periodic trajectory at a fixed altitude $H_u$ and with a given period $T$.
To make the problem tractable, the period $T$ is equally divided into $N$ time slots of duration $\delta  = T/N$. As a result, the trajectory of the UAV can be approximated by the $N$ two-dimensional sequences ${{\bf{q}}_u}\left[ n \right] = {\left[ {q_u^x\left[ n \right],q_u^y\left[ n \right]} \right]^T}$.
Note that the duration $\delta$ should be chosen to be  sufficiently small to satisfy ${V_{\max }}\delta  \ll {H_u}$, where $V_{\max}$ denotes the maximum UAV speed, so that the UAV's location can be considered as approximately unchanged within each time slot. The UAV mobility constraints are given below:
\begin{align}
&\left\| {{{\bf{q}}_u}\left[ n \right] - {{\bf{q}}_u}\left[ {n - 1} \right]} \right\| \le {V_{\max }}\delta ,\forall n, \label{Pconst5} \\
&{{\bf{q}}_u}\left[ 0 \right] = {{\bf{q}}_{\rm{I}}},{{\bf{q}}_u}\left[ N \right] = {{\bf{q}}_{\rm{F}}}, \label{Pconst6}
\end{align}
where  ${{\bf{q}}_{\rm{I}}}$ and ${{\bf{q}}_{\rm{F}}}$ represent the UAV's initial and final location, respectively.

It is assumed that the IRS has $M$ reflecting elements, and the reflection coefficient matrix of IRS $k$ at time slot $n$ is defined by the diagonal matrix  ${{\bf{\Theta }}_k}\left[ {n} \right]{\rm{ = diag}}\left( {{e^{j{\theta _{k,1}}\left[ {n} \right]}}, \ldots ,{e^{j{\theta _{k,M}}\left[ {n} \right]}}} \right)$, where $\theta_{k,m}[n]$ denotes the phase shift corresponding to the $m$th  reflecting element of IRS $k$ at time slot $n$ \cite{wu2019intelligent}, \cite{wu2019weighted},\cite{zhou2020Intelligent},\cite{zhang2020Capacity}. Let  ${\bf{h}}_{1,k}[n] \in {{\mathbb  C}^{ M \times 1 }}$, ${\bf{h}}_{2,k}[n] \in {{\mathbb  C}^{M \times 1}}$, and $h_3[n]\in {\mathbb  C}^{1 \times 1}$  respectively denote the complex equivalent baseband  channel vector between the UAV and the  $k$th IRS, between the $k$th IRS and the BS, and  between the UAV and the BS, at time slot $n$, $\forall k \in {\cal K} $. To capture both the large-scale and small-scale fading, we model  all channels as Rician \cite{zhan2020aerial}. Specifically, the channel coefficient between the UAV and IRS $k$ at time slot $n$ is given by  \cite{fundamentals2005Tse}
\begin{align}
{{\bf{h}}_{1,k}}\left[ n \right] = \sqrt {{\beta _{1,k}}\left[ n \right]} \left( {\sqrt {{{{K_1}} \over {{K_1} + 1}}} {\bf{h}}_{1,k}^{{\rm{LoS}}}\left[ n \right] + \sqrt {{1 \over {{K_1} + 1}}} {\bf{h}}_{1,k}^{{\rm{NLoS}}}\left[ n \right]} \right),
\end{align}
where ${{\beta _{1,k}}\left[ n \right]}$  represents the large-scale fading channel coefficient, ${{\bf{h}}_{1,k}^{{\rm{LoS}}}\left[ n \right]}$ and ${{\bf{h}}_{1,k}^{{\rm{NLoS}}}\left[ n \right]}$  denote the deterministic line-of-sight (LoS) channel component and the small-scale fading component, respectively, and $K_1$ is the Rician factor. The value of ${{\beta _{1,k}}\left[ n \right]}$ is related to the communication  distance, and is given by
\begin{align}
{\beta _{1,k}}\left[ n \right] = {{{\beta _0}} \over {d_{1,k}^{{\alpha _1}}\left[ n \right]}} = {{{\beta _0}} \over {{{\left( {{{\left\| {{{\bf{q}}_u}\left[ n \right] - {{\bf{q}}_{s,k}}} \right\|}^2} + {{\left( {{H_u} - {H_s}} \right)}^2}} \right)}^{{\alpha _1}/2}}}}, \label{deterinistic_beta1}
\end{align}
where ${{\beta _0}}$ denotes the channel power at the reference distance of $1$ meter, ${d_{1,k}}\left[ n \right]$ is the distance between the UAV and IRS $k$, and ${{\alpha _1}}$ denotes the path loss exponent. We assume that the IRS employs a uniform linear  array (ULA) of reflecting elements, and thus ${\bf{h}}_{1,k}^{{\rm{LoS}}}\left[ n \right]$ is given by \cite{fundamentals2005Tse}

\begin{align}
{\bf{h}}_{1,k}^{{\rm{LoS}}}\left[ n \right]  ={e^{ - j{{2\pi {d_{1,k}}\left[ n \right]} \over \lambda }}} \times{\left[ {1,{e^{ - j{{2\pi d} \over \lambda }\cos {\phi _{1,k}}\left[ n \right]}}, \ldots {e^{ - j{{2\pi \left( {M - 1} \right)d} \over \lambda }\cos {\phi _{1,k}}\left[ n \right]}}} \right]^T},
\end{align}
where $d$ denotes the IRS element spacing, $\lambda$ denotes the carrier wavelength, and $\cos {\phi _{1,k}}\left[ n \right] = {{ q_{s,k}^x-q_u^x\left[ n \right]} \over {{d_{1,k}}\left[ n \right]}}$ is the cosine of the angle of arrival (AoA) \cite{li2020Reconfigurable},\cite{hui2019Reflections}. The elements of ${{\bf{h}}_{1,k}^{{\rm{NLoS}}}\left[ n \right]}$ of the non-LoS component are assumed to be independent and identically distributed  and follow circularly symmetric complex Gaussian distribution with zero mean and unit variance.

Similarly, the channel vector between IRS $k$ and the BS at time slot $n$ is expressed as
\begin{align}
{{\bf{h}}_{2,k}}\left[ n \right] =
\sqrt {{\beta _{2,k}}} \left( {\sqrt {{{{K_2}} \over {{K_2} + 1}}} {\bf{h}}_{2,k}^{{\rm{LoS}}} + \sqrt {{1 \over {{K_2} + 1}}} {\bf{h}}_{2,k}^{{\rm{NLoS}}}\left[ n \right]} \right),
\end{align}
where ${\beta _{2,k}} = {{{\beta _0}} \over {d_{2,k}^{{\alpha _2}}}}$, ${d_{2,k}} = \sqrt {{{\left\| {{{\bf{q}}_b} - {{\bf{q}}_{s,k}}} \right\|}^2} + {{\left( {{H_b} - {H_s}} \right)}^2}} $, ${{\alpha _2}}$ represents the path loss exponent, and  $K_2$ is the corresponding Rician factor.  ${\bf{h}}_{2,k}^{{\rm{LoS}}} = {e^{  -j{{2\pi {d_{2,k}}} \over \lambda }}}{\left[ {1,{e^{ - j{{2\pi d} \over \lambda }\cos {\phi _{2,k}}}}, \ldots {e^{ - j{{2\pi \left( {M - 1} \right)d} \over \lambda }\cos {\phi _{2,k}}}}} \right]^T}$, where $\cos {\phi _{2,k}} = {{q_b^x - q_{s,k}^x} \over {{d_{2,k}}}}$ denotes the cosine of the angle of departure (AoD). The elements of ${{\bf{h}}_{2,k}^{{\rm{NLoS}}}\left[ n \right]}$ are also assumed to be independent and identically distributed and follow circularly symmetric complex Gaussian distribution with zero mean and unit variance.

Finally, for the UAV-BS link at time slot $n$ we have
\begin{align}
{{{h}}_3}\left[ n \right] = \sqrt {{\beta _3}\left[ n \right]} \left( {\sqrt {{{{K_3}} \over {{K_3} + 1}}} {{h}}_3^{{\rm{LoS}}}\left[ n \right] + \sqrt {{1 \over {{K_3} + 1}}} {{h}}_3^{{\rm{NLoS}}}\left[ n \right]} \right),
\end{align}
where ${\beta _3}\left[ n \right] = {{{\beta _0}} \over {d_3^{{\alpha _3}}\left[ n \right]}}$, ${d_3}\left[ n \right] = \sqrt {{{\left\| {{{\bf{q}}_u}\left[ n \right] - {{\bf{q}}_b}} \right\|}^2} + {{\left( {{H_b} - {H_u}} \right)}^2}} $, ${{h}}_3^{{\rm{LoS}}}\left[ n \right] = {e^{ - j{{2\pi {d_3}\left[ n \right]} \over \lambda }}}$, path-loss exponent ${{\alpha _3}}$, and Rician factor $K_3$. In addition, ${{{h}}_{3}^{{\rm{NLoS}}}\left[ n \right]} \sim {\cal CN}\left( {{{0 }},{{1 }}} \right)$.

\textbf{\emph{Remark 1:}} Although in this paper we  adopt  the ULA at IRS to facilitate the purpose of exposition, all the proposed algorithms are applicable to the case of    uniform planar  array (UPA) adopted at IRS with only a slight modifications for optimizing  IRS phase shift discussed in  \eqref{theorem2} in  Section III.

\textbf{\emph{Remark 2:}} Generally, there are two main approaches for the IRS-involved channel acquisition, depending on whether the IRS elements are equipped with receive radio frequency (RF) chains or not \cite{WU2020towards}. For the first approach with active receive RF chains, conventional channel estimation methods can be applied for the IRS to estimate the channels of the UAV-IRS and IRS-BS links, respectively. In contrast, for the second approach without receive RF chains at the IRS, the IRS reflection patterns can be designed jointly with the uplink pilots to estimate the concatenated UAV-IRS-BS channel and UAV-BS channel  e.g., \cite{lin2020adaptive}, \cite{zheng2020Intelligent}. 

Typically, the symbol rate for the IRS transmission is much lower than that for the primary (UAV) transmission due to the limited computational and communication capabilities at the IRS. 
To describe  it clearly, the frame structure for the IRS symbol, primary symbol, and channel coherence time  is shown in Fig.~\ref{fig2}. We assume that the duration of each UAV time slot equals the channel coherence time, i.e, $\delta {\rm{ = }}{T_c}$. In the figure, $x_{s,k}[n,n_1]$ represents the  $k$th IRS symbol transmitted to the BS in the  $n_1$th block of  time slot $n$, and  $x_u[n,n_1,l]$ is the primary symbol  transmitted from the UAV to the BS at the $l$th sub-block of block $n_1$ within time slot $n$.
Denote by $T_s$ and $T_u$ the durations of each IRS symbol and primary symbol, respectively.  Without loss of generality, we assume that each IRS symbol covers $  L$ primary symbols, namely $T_s=  LT_u$, where $  L$ is an integer, and $  L \gg 1$. In addition, we assume  $\delta =T_c=N_1T_s$, where $N_1$ is an integer, and $N_1 \gg 1$.

To facilitate the system design, we consider a widely used wake-up communication scheduling approach \cite{hua2019energy},\cite{wu2018Joint}, where the UAV can only communicate with at most one IRS\footnote{Strictly speaking,  the UAV directly communicates with the controller at IRS $k$ rather than IRS itself. In the sequel,  we will use the two terminologies interchangeably.} at any time slot $n$. Define the scheduling variable $a_k[n]$, where $a_k[n]=1$ indicates that IRS $k$ is served by the UAV, and $a_k[n]=0$ otherwise. We then have the following scheduling constraints
\begin{align}
&\sum\limits_{k = 1}^K {{a_k}\left[ n \right]}  \le 1,\forall k,n, \label{Pconst2}\\
&{a_k}\left[ n \right] \in \left\{ {0,1} \right\},\forall k, n. \label{Pconst3}
\end{align}
If  IRS $k$  is communicating with the UAV in time slot $n$, the signal received by the BS at the $l$th sub-block of block $n_1$ within time slot $n$ is given by
\begin{align}
{y_{r,k}}\left[ {n,n_1,l} \right] =& \underbrace {\sqrt P \left( {{\bf h}_{2,k}^H\left[ {n} \right]{{\bf \Theta} _k}\left[ {n} \right]{{\bf h}_{1,k}}\left[ {n,} \right]{x_{s,k}}\left[ n,n_1 \right]} \right){x_u}\left[ {n,n_1,l} \right]}_{{\rm{IRS \text{-} aided~ link}}}
{\rm{ + }}\underbrace {\sqrt P {h_3}\left[ {n} \right]{x_u}\left[ {n,n_1,l} \right]}_{{\rm{direct~link}}} \notag\\
& +w\left[ {n,n_1,l} \right],\label{sectionIIreceivedsignal}
\end{align}
where $P$ denotes the transmit  power at the UAV,  ${{x_u[n,n_1,l]}} \sim {\cal CN}\left( {0,1} \right)$,  ${{w[n,n_1,l]}} \sim {\cal CN}\left( {{{0}},{\sigma ^2}} \right)$ denotes the noise at IRS $k$ with power $\sigma^2$. We adopt the simple but widely used  on-off keying (OOK) modulation for IRS's information transmission, i.e., ${x_{s,k}}\left[ n,n_1 \right] = \left\{ {0,1} \right\}$. 

Since the IRS's own signal and UAV's primary signal
	are simultaneously received by the BS, to detect the composite signals correctly, several  detectors  such as maximum-likelihood (ML) detector, linear detector, and successive interference cancellation (SIC)-based detector, can be applied \cite{yang2018cooperative}. Furthermore, the strength of the signal received from UAV is generally much larger  than that received from the IRS due to the following two reasons.
	First, the direct link between the UAV and the BS is always dominated by LoS due  to the less scatters in the sky. Second, since the IRS consists of a large number of reflecting elements, it would significantly enhances the UAV's signal transmission via adjusting the phase shifters. In addition, 
	the SIC receiver decodes the stronger signal first, subtracts it from the composite signal, and extracts the weaker one from the residue.  Therefore, the SIC based detector is practically  appealing and   applied in this paper   \cite{yang2018cooperative}, \cite{verdu1998multiuser}.
%	Specifically, we first decode the primary signal (UAV transmission signal) by treating the IRS's own signal as interference. After correctly decoding the primary signal,  subtracting the primary signal  from the composite signal, and then decode the IRS's  signal.

Define   $h\left[ n \right] = {h_3}\left[ n \right] + {\bf{h}}_{2,k}^H\left[ n \right]{{\bf{\Theta }}_k}\left[ n \right]{{\bf{h}}_{1,k}}\left[ n \right]{x_{s,k}}\left[ {n,{n_1}} \right]$.
 It can  be observed that  $h\left[ n \right] $ contains ${{x_{s,k}}\left[ {n,{n_1}} \right]}$, which changes relatively fast as compared to the channel variation. In other words, the IRS's reflected signal ${{x_{s,k}}\left[ {n,{n_1}} \right]}$ plays
the role of fast-varying channel components, making  channel  $h\left[ n \right]$ vary over the primary signal ${x_u}\left[ {n,{n_1},l} \right]$ shown in \eqref{sectionIIreceivedsignal}.
According to \cite{zhang2020large}, [\cite{fundamentals2005Tse}, Appendix B.7], \cite{long2020symbiotic},  the  achievable rate (bps/Hz) for the primary (UAV) system assisted by IRS $k$ is given by
\begin{align}
&{{\bar R}_{u,k}}\left[ {n,{n_1},l} \right] \notag\\
&= {{\mathbb E}_{h\left[ n \right]}}\left\{ {{{\log }_2}\left( {1 + \frac{{P{{\left| {{h_3}\left[ n \right] + {\bf{h}}_{2,k}^H\left[ n \right]{{\bf{\Theta }}_k}\left[ n \right]{{\bf{h}}_{1,k}}\left[ n \right]{x_{s,k}}\left[ {n,{n_1}} \right]} \right|}^2}}}{{{\sigma ^2}}}} \right)} \right\}\notag\\
&  \overset{(a)}{=}{\mathbb E}_{x_{s,k}[n,n_1]}\left\{ {\log _2}\left( {1 + {{P{{\left| {{h_3}\left[ {n} \right] + {\bf h}_{2,k}^H\left[ {n} \right]{{\bf \Theta} _k}\left[ {n} \right]{{\bf h}_{1,k}}\left[ {n} \right]{x_{s,k}}\left[ n,n_1 \right]} \right|}^2}} \over {{\sigma ^2}}}} \right)\right\}\notag\\
&\overset{(b)}{=} \rho {\log _2}\left( {1 + {{P{{\left| {{h_3}\left[ {n} \right] + {\bf h}_{2,k}^H\left[ {n} \right]{{\bf \Theta} _k}\left[ {n} \right]{{\bf h}_{1,k}}\left[ {n} \right]} \right|}^2}} \over {{\sigma ^2}}}} \right)+ \left( {1 - \rho } \right){\log _2}\left( {1 + {{P{{\left| {{h_3}\left[ {n} \right]} \right|}^2}} \over {{\sigma ^2}}}} \right).\label{sectionIIprimaryrate}
\end{align}
where $(a)$ holds since under a channel coherence time, ${{h_3}\left[ n \right]}$, ${{{\bf{h}}_{1,k}}\left[ n \right]}$, and ${{{\bf{h}}_{2,k}}\left[ n \right]}$ are all invariant, while $h\left[ n \right]$ varies with the IRS's reflected signal   ${{x_{s,k}}\left[ {n,{n_1}} \right]}$. Equality $(b)$ holds since we assume that the probability for sending symbol ``1'' at  IRS $k$ is $\rho$ $(0\le\rho\le1)$, and that for sending symbol ``0'' at  IRS $k$ is $1-\rho$,  $\forall k$. It can be seen that the primary rate for  each sub-block within time slot $n$ is the same. Thus, the  achievable  rate  for the primary system assisted by IRS $k$  at time slot $n$
is given by ${{\bar R}_{u,k}}\left[ {n} \right]={{\bar R}_{u,k}}\left[ {n,n_1,l} \right]$.

After correctly decoding the primary signal  $x_u[n,n_1,l]$,   subtracting the primary signal  from the composite signal, and we can obtain the intermediate signal as \cite{verdu1998multiuser}
%After decoding $x_u[n,n_1,l]$, successive interference cancellation (SIC) is  applied at the BS tohus,  we can obtain the intermediate signal as directly remove  the direct link interference from the composite signals \cite{verdu1998multiuser}. T
\begin{align}
{{\hat y}_{r,k}}\left[ {n,n_1,l} \right] =\sqrt P \left( {{\bf h}_{2,k}^H\left[ {n} \right]{{\bf \Theta} _k}\left[ {n} \right]{{\bf h}_{1,k}}\left[ {n} \right]{x_{s,k}}\left[ n,n_1\right]} \right){x_u}\left[ {n,n_1,l} \right]+ w\left[ {n,n_1,l} \right].
\end{align}
For the different IRS reflected   symbols, the  signals received at  the BS have different amplitude values as
\begin{align}
{{\tilde y}_{r,k}}\left[ {n,{n_1},l} \right] = \left\{ \begin{array}{l}
\sqrt P \left( {{\bf{h}}_{2,k}^H\left[ n \right]{{\bf{\Theta }}_k}\left[ n \right]{{\bf{h}}_{1,k}}\left[ n \right]} \right){x_u}\left[ {n,{n_1},l} \right] + w\left[ {n,{n_1},l} \right]{\rm{,}}{\kern 1pt} {\kern 1pt} {\kern 1pt} {\kern 1pt} {\kern 1pt} {\rm{for}}{\kern 1pt} {\kern 1pt} {\kern 1pt} {\kern 1pt} {\kern 1pt} {\kern 1pt} {\rm{symbol}}{\kern 1pt} {\kern 1pt} {\kern 1pt} {\kern 1pt} {\kern 1pt} 1,\\
w\left[ {n,{n_1},l} \right]{\rm{,}}{\kern 1pt} {\kern 1pt} {\kern 1pt} {\kern 1pt} {\kern 1pt} {\rm{for}}{\kern 1pt} {\kern 1pt} {\kern 1pt} {\kern 1pt} {\kern 1pt} {\kern 1pt} {\rm{symbol}}{\kern 1pt} {\kern 1pt} {\kern 1pt} {\kern 1pt} {\kern 1pt} 0.
\end{array} \right.
\end{align}
It is not difficult to check that{\footnote{Note that although $x_u[n,n_1,l]$ is known at the BS after decoding,  $x_u[n,n_1,l]$ still can be regarded as a random variable  following  circularly symmetric complex Gaussian distribution  during  each one  IRS symbol since  each IRS symbol covers $ L$ primary symbols.}}
\begin{align}
{{\tilde y}_{r,k}}\left[ {n,{n_1},l} \right] \sim \left\{ \begin{array}{l}
{\cal CN}\left( {0,P{{\left| {{\bf{h}}_{2,k}^H\left[ n \right]{{\bf{\Theta }}_k}\left[ n \right]{{\bf{h}}_{1,k}}\left[ n \right]} \right|}^2} + {\sigma ^2}} \right){\rm{,}}{\kern 1pt} {\kern 1pt} {\kern 1pt} {\kern 1pt} {\kern 1pt} {\rm{for}}{\kern 1pt} {\kern 1pt} {\kern 1pt} {\kern 1pt} {\kern 1pt} {\kern 1pt} {\rm{symbol}}{\kern 1pt} {\kern 1pt} {\kern 1pt} {\kern 1pt} {\kern 1pt} 1,\\
{\cal CN}\left( {0,{\sigma ^2}} \right){\rm{,}}{\kern 1pt} {\kern 1pt} {\kern 1pt} {\kern 1pt} {\kern 1pt} {\rm{for}}{\kern 1pt} {\kern 1pt} {\kern 1pt} {\kern 1pt} {\kern 1pt} {\kern 1pt} {\rm{symbol}}{\kern 1pt} {\kern 1pt} {\kern 1pt} {\kern 1pt} {\kern 1pt} 0,
\end{array} \right.
\end{align}
  We adopt  a simple joint-energy detector for detecting IRS's symbols \cite{qian2017Noncoherent}, \cite{hua2020Bistatic}. Define ${{\bar y}_{r,k}}\left[ {n,{n_1}} \right] = \sum\limits_{l = 1}^L {{{\left| {{{\tilde y}_{r,k}}\left[ {n,{n_1},l} \right]} \right|}^2}} $.
It can be readily checked that  a random variable ${{\bar y}_{r,k}}\left[ {n,{n_1}} \right]$ is the sum of $L$ independent identically distributed  central chi-squared random variables with two degrees of freedom. Suppose that the   symbol ``1'' hypothesis is  ${\cal H}_1$ and symbol ``0'' hypothesis is  ${\cal H}_0$, we can readily obtain the  expectation and variance of ${{{\left| {{{\tilde y}_{r,k}}\left[ {n,{n_1},l} \right]} \right|}^2}}$
as 
\begin{align}
&{\mathbb E}\left\{ {{{\left| {\tilde y_{r,k}\left[ n,n_1,l \right]} \right|}^2}} \right\} = \sigma _i^2, ~{\rm under} ~{\cal H}_i,  i=\{0,1\},\notag\\
&{\mathop{{{ \rm Var} }}} \left\{ {{{\left| {\tilde y_{r,k}\left[ n,n_1,l \right]} \right|}^2}} \right\} = \sigma _i^4, ~{\rm under} ~ {\cal H}_i, i=\{0,1\},
\end{align}
with $\sigma _1^2 = P{\left| {{\bf{h}}_{2,k}^H\left[ n \right]{{\bf{\Theta }}_k}\left[ n \right]{{\bf{h}}_{1,k}}\left[ n \right]} \right|^2} + {\sigma ^2}$ and $\sigma _0^2 = {\sigma ^2}$.
Based on the central limit theorem, when $L$ is large, the distribution of ${{\bar y}_{r,k}}\left[ {n,{n_1}} \right]$ asymptotically approaches
a Gaussian distribution as 
\begin{align}
{{\bar y}_{r,k}}[n,n_1] \sim {\cal N}\left( {L\sigma _i^2,L\sigma _i^4} \right), ~{\rm under} ~{\cal H}_i,  i=\{0,1\},
\end{align}
Define  probability density function $f\left( {\bar y_{r,k}[n,n_1]|{\cal H}_i} \right) = \frac{1}{{\sqrt {2\pi L} \sigma _i^2}}\exp \left( { - \frac{{{{\left( {\bar y_{r,k+}[n,n_1] - L\sigma _i^2} \right)}^2}}}{{2L\sigma _i^4}}} \right),i=\{0,1\}$. Following from \cite{kay1993fundamentals}, the decision criteria is if 
\begin{align}
f\left( {\bar y_{r,k}[n,n_1]|{{\cal H}_1}} \right) \ge f\left( {\bar y_{r,k}[n,n_1]|{{\cal H}_0}} \right), 
\end{align}
then symbol ``1'' is chosen,  otherwise, symbol ``0'' is chosen. As such, 
the BER of IRS  can be derived from 
\begin{align}
&{P_{e,k}}\left[ {n,{n_1}} \right] = \notag\\
&\Pr \left( {{{\cal H}_1}} \right)\Pr \left( {{{\bar y}_{r,k}}\left[ {n,{n_1}} \right] < \bar y_r^{th}\left[ {n,{n_1}} \right]|{{\cal H}_1}} \right) + \Pr \left( {{{\cal H}_0}} \right)\Pr \left( {{{\bar y}_r}\left[ {n,{n_1}} \right] \ge y_r^{th}\left[ {n,{n_1}} \right]|{{\cal H}_0}} \right),
\end{align}
where ${\bar y_r^{th}\left[ {n,{n_1}} \right]}$ denotes the decision  threshold. The optimal decision  threshold ${\bar y_r^{th,opt}\left[ {n,{n_1}} \right]}$ can be derived by taking the first derivative of ${P_{e,k}}\left[ {n,{n_1}} \right]$ with respect to (w.r.t.) ${\bar y_r^{th}\left[ {n,{n_1}} \right]}$, which is given by 
\begin{align}
{\bar y^{th,opt}_r} = \frac{{L\sigma _1^2\sigma _0^2}}{{\sigma _1^2 + \sigma _0^2}}\left[ {1 + \sqrt {1 + \frac{{2\left( {\sigma _1^2 + \sigma _0^2} \right)\ln \frac{{\sigma _1^2}}{{\sigma _0^2}}}}{{L\left( {\sigma _1^2 - \sigma _0^2} \right)}}} } \right]. \label{csr_new1}
\end{align}
Therefore, the BER for detecting IRS $k$'s reflected symbol can be derived as
\begin{align}
{P_{e,k}[n,n_1]} &= \rho Q\left( {\frac{{L\sigma _1^2 - \bar y^{th,opt}_r}}{{\sqrt L \sigma _1^2}}} \right) + (1-\rho)Q\left( {\frac{{\bar y^{th,opt}_r - L\sigma _0^2}}{{\sqrt L \sigma _0^2}}} \right), \notag\\
&\overset{(a)}{\approx}  Q\left( {\sqrt L \frac{{\sigma _1^2 - \sigma _0^2}}{{\sigma _1^2 + \sigma _0^2}}} \right) \notag\\
&{\rm{ = }}Q\left( {\sqrt L \frac{{P{{\left| {{\bf{h}}_{2,k}^H\left[ n \right]{{\bf{\Theta }}_k}\left[ n \right]{{\bf{h}}_{1,k}}\left[ n \right]} \right|}^2}}}{{P{{\left| {{\bf{h}}_{2,k}^H\left[ n \right]{{\bf{\Theta }}_k}\left[ n \right]{{\bf{h}}_{1,k}}\left[ n \right]} \right|}^2}{\rm{ + 2}}{\sigma ^2}}}} \right), \label{ber}
\end{align}
with  the $Q$ function given by $Q\left( x \right) = \frac{1}{{\sqrt {2\pi } }}\int_x^\infty  {{e^{ - \frac{{{t^2}}}{2}}}} dt$. $(a)$  holds since for a large value $L$, ${\bar y_r^{th,opt}}$ given in \eqref{csr_new1} approaches $\frac{{2L\sigma _1^2\sigma _0^2}}{{\sigma _1^2 + \sigma _0^2}}$. 
We can see that   the BER of IRS $k$ within  each sub-block of time slot $n$ is the same. Thus,  the  BER  for  IRS $k$  at time slot $n$
is given by ${P_{e,k}}\left[ n \right] = {P_{e,k}}\left[ {n,{n_1}} \right]$.

In this paper,  we are interested in the average  communication throughput and average BER.
\textbf{\emph{Theorem 1:}} The average achievable rate for the primary system, i.e., ${\mathbb E}\{{{\bar R}_{u,k}[n]}\}$, is upper bounded by
\begin{align}
{\mathbb E}\left\{ {{{\bar R}_{u,k}}\left[ {n} \right]} \right\} \le& { \hat R_{u,k}}\left[ {n} \right]\overset{\triangle}{\rm{ = }}\left( {1 - \rho } \right){\log _2}\left( {1 + \frac{{P{\beta _3}\left[ n \right]}}{{{\sigma ^2}}}} \right){\rm{ + }}\notag\\
&\rho {\log _2}\left( {1 + \frac{{P\left( {{{\left| {{x_{0,k}}\left[ n \right]} \right|}^2} + \frac{{\left( {{K_1} + {K_2}{\rm{ + }}1} \right)M{\beta _{1,k}}\left[ n \right]{\beta _{2,k}}}}{{\left( {{K_1} + 1} \right)\left( {{K_2} + 1} \right)}}{\rm{ + }}\frac{{{\beta _3}\left[ n \right]}}{{{K_3} + 1}}} \right)}}{{{\sigma ^2}}}} \right), \label{sectionIIprimaryrateNEW}
\end{align}
where ${x_{0,k}}\left[ n \right] = \sqrt {\frac{{{K_3}{\beta _3}\left[ n \right]}}{{{K_3} + 1}}} h_3^{{\rm{LoS}}}\left[ n \right] + \sqrt {\frac{{{K_1}{K_2}{\beta _{1,k}}\left[ n \right]{\beta _{2,k}}}}{{\left( {{K_1} + 1} \right)\left( {{K_2} + 1} \right)}}} {\left( {{\bf{h}}_{2,k}^{{\rm{LoS}}}\left[ n \right]} \right)^H}{{\bf{\Phi }}_k}\left[ n \right]{\bf{h}}_{1,k}^{{\rm{LoS}}}\left[ n \right]$.

\hspace*{\parindent}\textit{Proof}: Please refer to Appendix~\ref{appendix1}. \hfill\rule{2.7mm}{2.7mm}

In Theorem 1, we can see that $ {\hat R}_{u,k}[n]$  is determined by the deterministic LoS channel components $\left\{ {{h_3}^{\rm LoS}\left[ n \right],{{\bf{h}}_{1,k}^{\rm LoS}}\left[ n \right],{{\bf{h}}_{2,k}^{\rm LoS}}\left[ n \right]} \right\}$, the large-scale fading coefficients $\left\{ {{\beta _{1,k}}\left[ n \right],{\beta _{2,k}},{\beta _3}\left[ n \right]} \right\}$, and the IRS phase shift matrix ${{\bf{\Phi }}_k}\left[ n \right]$. It is worth pointing out that the above approximation will be tight if the SNR is sufficiently high \cite{Han2019large}.

Define SNR ${{\bar \gamma} _k}\left[ n \right] = \frac{{\left| {{\bf{h}}_{2,k}^H\left[ n \right]{{\bf{\Theta }}_k}\left[ n \right]{{\bf{h}}_{1,k}}\left[ n \right]} \right|}}{{{\sigma ^2}}}$. Similarly, we  can obtain the average SNR for IRS $k$ at time slot $n$ as 
\begin{align}
 \gamma_k[n]\overset{\triangle}{=}{\mathbb E}\left\{ {{{\bar \gamma }_k}\left[ n \right]} \right\}{\rm{ = }}\frac{{\left( {{{\left| {{{\bar x}_{0,k}}\left[ n \right]} \right|}^2} + \frac{{\left( {{K_1} + {K_2}{\rm{ + }}1} \right)M{\beta _{1,k}}\left[ n \right]{\beta _{2,k}}}}{{\left( {{K_1} + 1} \right)\left( {{K_2} + 1} \right)}}} \right)}}{{{\sigma ^2}}}, \label{sectionIIIRSrateNEW}
\end{align}
where ${{\bar x}_{0,k}}\left[ n \right] = \sqrt {\frac{{{K_1}{K_2}{\beta _{1,k}}\left[ n \right]{\beta _{2,k}}}}{{\left( {{K_1} + 1} \right)\left( {{K_2} + 1} \right)}}} {\left( {{\bf{h}}_{2,k}^{{\rm{LoS}}}\left[ n \right]} \right)^H}{{\bf{\Phi }}_k}\left[ n \right]{\bf{h}}_{1,k}^{{\rm{LoS}}}\left[ n \right]$.

%Similarly, we  can obtain an upper bound of the reflecting  rate for IRS $k$ at time slot $n$ as
%\begin{align}
%{\mathbb E}\left\{ {{{\bar R}_{s,k}}\left[ n \right]} \right\} \le { \hat R_{s,k}}\left[ n \right] = {\log _2}\left( {1 + \frac{{LP\left( {{{\left| {{{\bar x}_{0,k}}\left[ n \right]} \right|}^2} + \frac{{\left( {{K_1} + {K_2}{\rm{ + }}1} \right)M{\beta _{1,k}}\left[ n \right]{\beta _{2,k}}}}{{\left( {{K_1} + 1} \right)\left( {{K_2} + 1} \right)}}} \right)}}{{{\sigma ^2}}}} \right), \label{sectionIIIRSrateNEW}
%\end{align}
%where ${{\bar x}_{0,k}}\left[ n \right] = \sqrt {\frac{{{K_1}{K_2}{\beta _{1,k}}\left[ n \right]{\beta _{2,k}}}}{{\left( {{K_1} + 1} \right)\left( {{K_2} + 1} \right)}}} {\left( {{\bf{h}}_{2,k}^{{\rm{LoS}}}\left[ n \right]} \right)^H}{{\bf{\Phi }}_k}\left[ n \right]{\bf{h}}_{1,k}^{{\rm{LoS}}}\left[ n \right]$.
%In the sequel, the upper bounds  for the primary and reflecting rate  are used  as our utility functions. It is worth pointing out that the above approximations will be tight if the SNR is sufficiently high \cite{Han2019large}.

\subsection{Problem Formulation}
For the first   scenario, our goal is to minimize the  weighted sum BER  among all IRSs over all the time slots  by jointly optimizing the UAV trajectory, IRS phase shift matrix, and IRS scheduling. Accordingly, the  problem can be formulated as
\begin{subequations}  \label{wsrP}
	\begin{align}
	&\mathop {\min }\limits_{{\theta _{k,m}}\left[ n \right],{{\bf{q}}_u}\left[ n \right],{a_k}\left[ n \right]} \sum\limits_{k = 1}^K {{w_k}} \sum\limits_{n = 1}^N {{a_k}\left[ n \right]{\mathbb E}\left\{ {{P_{e,k}}\left[ n \right]} \right\}}  \label{wsrPobj0}\\
	&{\rm{s}}{\rm{.t}}{\rm{.}}~\sum\limits_{k = 1}^K {{a_k}\left[ n \right]} {\hat R_{u,k}}{\kern 1pt} \left[ n \right]{\kern 1pt}  \ge {R_{{\rm{th}}}},\forall n,\label{wsrPconst1}\\
	&\qquad{\kern 1pt} 0 \le {\theta _{k,m}}[n] \le 2\pi ,\forall m,k,n,\label{wsrPconst4}\\
	&\qquad \eqref{Pconst5}, \eqref{Pconst6}, \eqref{Pconst2}, \eqref{Pconst3},
	\end{align}
\end{subequations}
where $w_k$ denotes the weighting factor for IRS $k$, with a higher value representing  a higher priority over other IRSs, and  ${R_{{\rm{th}}}}$ is the minimum rate requirement of the primary transmission system for any time slot $n$. Problem \eqref{wsrP} is challenging to solve mainly due to the following three reasons. First, the optimization variables $a_k[n]$ for communication
scheduling  are binary and thus \eqref{Pconst3}, \eqref{wsrPobj0}, and \eqref{wsrPconst1} involve integer constraints. Second, the IRS phase shift matrix, UAV trajectory, and IRS scheduling are intricately coupled in \eqref{wsrPobj0} and  \eqref{wsrPconst1}, which makes the problem non-convex. Third, the expression of ${\mathbb E}\{P_{e,k}[n]\}$ in the objective function is implicit w.r.t. the optimization variables. In general, there is no efficient method to optimally solve problem~\eqref{wsrP}.

% To address this issue,  a low-complexity algorithm based on the relaxation method is proposed to solve  problem~\eqref{wsrP}\cite{wu2018Joint},\cite{HUA2020throughput},\cite{Hua2018power}. In Section III, we prove that the  relaxed  solutions $a_k[n]$ obtained by solving  the proposed algorithm are binary, which means that no reconstruct strategy is needed, and thus the primary rate requirements in~\eqref{wsrPconst1} are always satisfied.

For the second  scenario, our goal is to minimize the maximum
BER among all IRS over all the time slots  by jointly optimizing the UAV trajectory, the IRS phase shift matrix, and the IRS scheduling. Accordingly, the  problem can be formulated as
\begin{subequations}  \label{P}
\begin{align}
&\mathop {\min }\limits_{{\theta _{k,m}}\left[ n \right],{{\bf{q}}_u}\left[ n \right],{a_k}\left[ n \right],R} R \label{Pobj0}\\
&{\rm{s}}.{\rm{t}}.~\frac{1}{N}\sum\limits_{n = 1}^N {{a_k}\left[ n \right]{\mathbb E}\left\{ {{P_{e,k}}\left[ n \right]} \right\}}  \le R,\forall k,\label{Pobj1}\\
&\qquad\sum\limits_{k = 1}^K {{a_k}\left[ n \right]} {\hat R_{u,k}}{\kern 1pt} \left[ n \right]{\kern 1pt}  \ge {R_{{\rm{th}}}},\forall n,\label{Pconst1}\\
&\qquad \eqref{Pconst5}, \eqref{Pconst6}, \eqref{Pconst2}, \eqref{Pconst3}, \eqref{wsrPconst4}.
\end{align}
\end{subequations}
The  left hand side of \eqref{Pobj1} denotes the  average BER of IRS $k$ over all $N$ time slots. Problem~\eqref{P} is still difficult to solve due to the similar challenges for the weighted sum BER problem discussed above.

%Although problem~\eqref{P} is similar to problem~\eqref{wsrP},  the relaxation-based  method cannot be applied to  problem~\eqref{P} (We will clarify  this issue in Section IV). 
%To address this issue and obtain a high-quality suboptimal solution, we first transform the binary constraint~\eqref{Pconst3} into  a series of equivalent equality constraints. Then, a novel  penalty-based algorithm is proposed to solve~\eqref{P} whose solution for $a_k[n]$ is guaranteed to be binary, and the constraints in~\eqref{Pconst1} are automatically satisfied.

\textbf{\emph{Remark 3:}} It can be seen that ${P_{e,k}}\left[ {n} \right]$ given in \eqref{ber} is a complicated  expression with an integral, which is challenging to analyze  directly. In addition,  it  can be readily verified that $Q$ function is monotonically decreasing with SNR ${{{\bar \gamma }_k}\left[ n \right]}$. Instead of maximizing the SNR directly,  we introduce a utility function ${{\cal F}\left( {{{\bar \gamma }_k}\left[ n \right]} \right)}$, which is a differential, concave and monotonically  increasing function w.r.t. ${{\bar \gamma} _k}\left[ n \right]$, to replace BER ${P_{e,k}}\left[ {n} \right]$. In addition, to obtain the average BER ${\mathbb E}\left\{ {{P_{e,k}}\left[ n \right]} \right\}$, we   set the upper bound of ${\mathbb E}\left\{ {{\cal F}\left( {{{\bar \gamma }_k}\left[ n \right]} \right)} \right\}$, i.e.,  ${\cal F}\left( {{\mathbb E}\left\{ {{{\bar \gamma }_k}\left[ n \right]} \right\}} \right)$,  as our design metric  for facilitating  the algorithm design. In the subsequent sections, instead of minimizing the BER directly, we aim to maximize the corresponding utility functions.
\section{Relaxation-based Algorithm for Weighted Sum BER Optimization Problem}
In this section, we propose a relaxation-based algorithm to solve problem~\eqref{wsrP}. Specifically, we  first   relax the binary scheduling variables into continuous variables, and divide the relaxed non-convex problem into two sub-problems, then  solve   these two sub-problems. However, even with this decomposition, the problem is still difficult to handle due to the non-convex cosine in both the objective function and constraints. To address this issue, we first obtain a closed-form solution for the IRS phase shift matrix for a given UAV trajectory and communication scheduling, and then substitute this expression into the original problem resulting in a joint IRS scheduling and UAV trajectory  optimization problem. We first develop the following theorem:

\textbf{\emph{Theorem 2:}} For any given UAV trajectory and IRS scheduling, the optimal IRS phase shift matrix that maximizes the primary rate and SNR  is given by
\begin{align}
{\theta^{\rm opt} _{k,m}}\left[ n \right] = {\rm{ - }}\frac{{2\pi d\left( {\left( {\cos {\phi _{2,k}} - \cos {\phi _{1,k}}\left[ n \right]} \right)\left( {m - 1} \right){\rm{ - }}\left( {{d_{1,k}}\left[ n \right] - {d_{2,k}}\left[ n \right]} \right){\rm{ + }}{d_3}\left[ n \right]} \right)}}{\lambda },\forall k,m,n. \label{theorem2}
\end{align}
\hspace*{\parindent}\textit{Proof}: Please refer to Appendix~\ref{appendix2}.\hfill\rule{2.7mm}{2.7mm}

From Appendix~\ref{appendix2}, we can see that the maximizer of the terms $\left| {{x_{0,k}}\left[ n \right]} \right|^2$ and $\left| {{\bar x_{0,k}}\left[ n \right]} \right|^2$ are respectively given by
\begin{align}
{\left| {{x_{0,k}^{ *}}\left[ n \right]} \right|^2}{\rm{ = }}\frac{{{K_3}{\beta _3}\left[ n \right]}}{{{K_3} + 1}}{\rm{ + }}\frac{{{K_1}{K_2}{M^2}{\beta _{1,k}}\left[ n \right]{\beta _{2,k}}}}{{\left( {{K_1} + 1} \right)\left( {{K_2} + 1} \right)}}{\rm{ + }}2M\sqrt {\frac{{{K_1}{K_2}{K_3}{\beta _{1,k}}\left[ n \right]{\beta _{2,k}}{\beta _3}\left[ n \right]}}{{\left( {{K_1} + 1} \right)\left( {{K_2} + 1} \right)\left( {{K_3} + 1} \right)}}},\label{SectionIII_1}
\end{align}
and
\begin{align}
{\left| {{{\bar x}_{0,k}^{*}}\left[ n \right]} \right|^2}{\rm{ = }}\frac{{{K_1}{K_2}{M^2}{\beta _{1,k}}\left[ n \right]{\beta _{2,k}}}}{{\left( {{K_1} + 1} \right)\left( {{K_2} + 1} \right)}}. \label{SectionIII_2}
\end{align}
Substituting \eqref{SectionIII_1}  and \eqref{SectionIII_2} in \eqref{sectionIIprimaryrateNEW} and \eqref{sectionIIIRSrateNEW}, respectively,  we   have
\begin{align}
{R_{u,k}}\left[ n \right] &= \left( {1 - \rho } \right){\log _2}\left( {1 + \frac{{P{\beta _3}\left[ n \right]}}{{{\sigma ^2}}}} \right)\notag\\
&{\rm{ + }}\rho {\log _2}\left( {1 + \frac{{P\left( {\left( {{{\rm{c}}_{k,1}}{\rm{ + }}{{\rm{c}}_{k,3}}} \right){\beta _{1,k}}\left[ n \right]{\rm{ + }}{{\rm{c}}_{k,2}}\sqrt {{\beta _{1,k}}\left[ n \right]{\beta _3}\left[ n \right]} {\rm{ + }}{\beta _3}\left[ n \right]} \right)}}{{{\sigma ^2}}}} \right), \label{SectionIII_3}
\end{align}
and
\begin{align}
{\gamma _k}\left[ n \right] = \frac{{\left( {{{\rm{c}}_{k,1}}{\rm{ + }}{{\rm{c}}_{k,3}}} \right){\beta _{1,k}}\left[ n \right]}}{{{\sigma ^2}}}, \label{SectionIII_4}
\end{align}
where ${{\rm{c}}_{k,1}} = \frac{{{K_1}{K_2}{M^2}{\beta _{2,k}}}}{{\left( {{K_1} + 1} \right)\left( {{K_2} + 1} \right)}}$, ${{\rm{c}}_{k,2}} = 2M\sqrt {\frac{{{K_1}{K_2}{K_3}{\beta _{2,k}}}}{{\left( {{K_1} + 1} \right)\left( {{K_2} + 1} \right)\left( {{K_3} + 1} \right)}}} $, and ${{\rm{c}}_{k,3}} = \frac{{\left( {1{\rm{ + }}{K_1}{\rm{ + }}{K_2}} \right)M{\beta _{2,k}}}}{{\left( {{K_1} + 1} \right)\left( {{K_2} + 1} \right)}}$.

As a result, the weighted sum BER optimization problem  \eqref{wsrP} can be  simplified as
\begin{subequations}  \label{wsrP1}
	\begin{align}
	&\mathop {\max }\limits_{{{\bf{q}}_u}\left[ n \right],{a_k}\left[ n \right]} \sum\limits_{k = 1}^K {{w_k}\sum\limits_{n = 1}^N {{a_k}\left[ n \right]{\cal F}\left( {\frac{{\left( {{{\rm{c}}_{k,1}}{\rm{ + }}{{\rm{c}}_{k,3}}} \right){\beta _{1,k}}\left[ n \right]}}{{{\sigma ^2}}}} \right)} }    \\
	&{\rm{s}}{\rm{.t}}{\rm{.}}~\sum\limits_{k = 1}^K {{a_k}\left[ n \right]} \left( {\rho {{\log }_2}\left( {1 + \frac{{P\left( {\left( {{{\rm{c}}_{k,1}}{\rm{ + }}{{\rm{c}}_{k,3}}} \right){\beta _{1,k}}\left[ n \right]{\rm{ + }}{{\rm{c}}_{k,2}}\sqrt {{\beta _{1,k}}\left[ n \right]{\beta _3}\left[ n \right]} {\rm{ + }}{\beta _3}\left[ n \right]} \right)}}{{{\sigma ^2}}}} \right)} \right.\notag\\
	&\qquad +{\left( {1 - \rho } \right){{\log }_2}\left( {1 + \frac{{P{\beta _3}\left[ n \right]}}{{{\sigma ^2}}}} \right)} \Bigg) \ge {R_{{\rm{th}}}},\forall n,\label{wsrPconst1NEW}\\
	&\qquad \eqref{Pconst5},\eqref{Pconst6},\eqref{Pconst2},\eqref{Pconst3}.
	\end{align}
\end{subequations}
It can be seen that  \eqref{wsrP1} only involves two variables, ${{{\bf{q}}_u}\left[ n \right]}$ and  ${{a_k}\left[ n \right]}$, and the cosine function no longer appears,  which thus make the problem easier to solve.  In the following, a low complexity algorithm based on the relaxation method is proposed. Note that different from \cite{wu2018Joint},\cite{HUA2020throughput},\cite{Hua2018power}, where the resulting continuous scheduling variables  need to be converted into binary. However, for problem \eqref{wsrP1}, we   prove that the converged relaxation scheduling variables are binary, which means that no reconstruct strategy is needed, and thus the UAV rate constraints \eqref{wsrPconst1NEW} are automatically satisfied. Specifically, we first relax the binary variable $a_k[n]$ to a continuous
variable, and rewrite constraint \eqref{Pconst3} as follows \cite{wu2018Joint},\cite{HUA2020throughput},\cite{Hua2018power}:
\begin{align}
0 \le {a_k}\left[ n \right] \le 1,\forall k,n. \label{Pconst3NEW}
\end{align}
We then decompose the relaxed problem into two separate subproblems, IRS scheduling and UAV trajectory optimization, and then alternately optimize each one.

\subsection{IRS Scheduling Optimization}
For any given UAV trajectory ${\bf q}_u[n]$, the IRS scheduling problem of \eqref{wsrP1} becomes
\begin{subequations}  \label{wsrP1-1}
	\begin{align}
	&\mathop {\max }\limits_{{a_k}\left[ n \right]} \sum\limits_{k = 1}^K {{w_k}\sum\limits_{n = 1}^N {{a_k}\left[ n \right]{\cal F}\left( {\frac{{\left( {{{\rm{c}}_{k,1}}{\rm{ + }}{{\rm{c}}_{k,3}}} \right){\beta _{1,k}}\left[ n \right]}}{{{\sigma ^2}}}} \right)} }  \label{wsrPobj0NEW}\\
	&{\rm{s}}{\rm{.t}}{\rm{.}}~\eqref{Pconst2},\eqref{wsrPconst1NEW},\eqref{Pconst3NEW}.
	\end{align}
\end{subequations}
Since both the objective function and constraints are linear w.r.t. $a_k[n]$, problem \eqref{wsrP1-1} is thus a linear optimization problem.

\textbf{\emph{Theorem 3:}} The optimal solution $a^{\rm opt}_k[n]$ to problem \eqref{wsrP1-1} is binary, i.e., $a^{\rm opt}_k[n] \in \{0,1\}$.

\hspace*{\parindent}\textit{Proof}: Please refer to Appendix~\ref{appendix3}.\hfill\rule{2.7mm}{2.7mm}

Theorem 3 shows that even though the binary constraint in the IRS scheduling problem of \eqref{wsrP1-1} has been relaxed, the obtained solution  is still a binary result. As such, no reconstruction operation is needed. In addition, since \eqref{wsrP1-1} is a linear optimization problem, it has very low computational complexity \cite{gondzio1996computational}.

\subsection{UAV Trajectory  Optimization}
For any given IRS scheduling $a_k[n]$, the UAV trajectory optimization problem of \eqref{wsrP1} becomes
\begin{subequations}  \label{wsrP2}
	\begin{align}
	&\mathop {\max }\limits_{{{\bf{q}}_u}\left[ n \right]} \sum\limits_{k = 1}^K {{w_k}\sum\limits_{n = 1}^N {{a_k}\left[ n \right]{\cal F}\left( {\frac{{\left( {{{\rm{c}}_{k,1}}{\rm{ + }}{{\rm{c}}_{k,3}}} \right){\beta _{1,k}}\left[ n \right]}}{{{\sigma ^2}}}} \right)} }    \label{wsrP2obj0NEW}\\
	%	&{\rm{s}}{\rm{.t}}{\rm{.}}~\sum\limits_{k = 1}^K {{a_k}\left[ n \right]} \left( {\rho {{\log }_2}\left( {1 + \frac{{P\left( {\left( {{{\rm{c}}_{k,1}}{\rm{ + }}{{\rm{c}}_{k,3}}} \right){\beta _{1,k}}\left[ n \right]{\rm{ + }}{{\rm{c}}_{k,2}}\sqrt {{\beta _{1,k}}\left[ n \right]{\beta _3}\left[ n \right]} {\rm{ + }}{\beta _3}\left[ n \right]} \right)}}{{{\sigma ^2}}}} \right)} \right.\notag\\
	%	&\qquad +{\left( {1 - \rho } \right){{\log }_2}\left( {1 + \frac{{P{\beta _3}\left[ n \right]}}{{{\sigma ^2}}}} \right)} \Bigg) \ge {R_{{\rm{th}}}},\forall n,\label{wsrP2const1NEW}\\
	&{\rm s.t.}~\eqref{Pconst5},\eqref{Pconst6},\eqref{wsrPconst1NEW}.
	\end{align}
\end{subequations}
Note that \eqref{wsrP2} is neither concave or quasi-concave due to the non-convex   constraints \eqref{wsrPconst1NEW} and  non-convex objective function \eqref{wsrP2obj0NEW}. In general, there is no efficient method
to obtain the optimal solution. In the following, we adopt the
successive convex optimization technique to solve \eqref{wsrP2}. To this end, we introduce additional slack variables $\{z_{1,k}[n]\}$ and $\{z_3[n]\}$, and recast  \eqref{wsrP2} as
\begin{subequations} \label{wsrP2-31}
	\begin{align}
	&\mathop {\max }\limits_{{{\bf{q}}_u}\left[ n \right],{z_{1,k}}\left[ n \right],{z_{3,k}}\left[ n \right]} \sum\limits_{k = 1}^K {{w_k}\sum\limits_{n = 1}^N {{a_k}\left[ n \right]{\cal F}\left( {\frac{{\left( {{{\rm{c}}_{k,1}}{\rm{ + }}{{\rm{c}}_{k,3}}} \right){z_{1,k}}\left[ n \right]}}{{{\sigma ^2}}}} \right)} } \label{wsrP2-31const1}\\
	&\qquad\sum\limits_{k = 1}^K {{a_k}\left[ n \right]} \left( {\rho {{\log }_2}\left( {1 + \frac{{P\left( {\left( {{{\rm{c}}_{k,1}}{\rm{ + }}{{\rm{c}}_{k,3}}} \right){z _{1,k}}\left[ n \right]{\rm{ + }}{{\rm{c}}_{k,2}}\sqrt {{z _{1,k}}\left[ n \right]{z _3}\left[ n \right]} {\rm{ + }}{z _3}\left[ n \right]} \right)}}{{{\sigma ^2}}}} \right)} \right.\notag\\
	&\qquad +{\left( {1 - \rho } \right){{\log }_2}\left( {1 + \frac{{P{z_3}\left[ n \right]}}{{{\sigma ^2}}}} \right)} \Bigg) \ge {R_{{\rm{th}}}},\forall n,\label{wsrP2-31const2}\\
	&\qquad{\beta _{1,k}}\left[ n \right] \ge {z_{1,k}}\left[ n \right],\forall k,n,\label{wsrP2-31const3}\\
	&\qquad{\beta _3}\left[ n \right] \ge {z_3}\left[ n \right],\forall n,\label{wsrP2-31const4}\\
	&\qquad \eqref{Pconst5}, \eqref{Pconst6}.
	\end{align}
\end{subequations}
It can be shown that at the optimal solution to \eqref{wsrP2-31}, we
must have ${\beta _{1,k}}\left[ n \right]= {z_{1,k}}\left[ n \right]$ and ${\beta _3}\left[ n \right]= {z_3}\left[ n \right],\forall k,n$, since otherwise we can always increase $z_{1,k}[n]$  (or $z_{3}[n]$) without decreasing the value of the objective. Therefore, problem \eqref{wsrP2-31} is equivalent to problem \eqref{wsrP2}. With this reformulation, objective function  \eqref{wsrP2-31const1} is now
concave w.r.t. $z_{1,k}[n]$, but with the new non-convex
constraints \eqref{wsrP2-31const3} and \eqref{wsrP2-31const4}.  The key observation is that in \eqref{wsrP2-31const3}, although ${\beta _{1,k}}\left[ n \right]$, defined in \eqref{deterinistic_beta1}, is not convex  w.r.t. ${\bf q}_u[n]$, it is convex  w.r.t. ${{{\left\| {{{\bf{q}}_u}\left[ n \right] - {{\bf{q}}_{s,k}}} \right\|}^2}}$. Recall that any convex
function is globally lower-bounded by its first-order Taylor
expansion at any feasible point \cite{boyd2004convex}. Therefore, for any local point ${{{\left\| {{\bf{q}}_u^r\left[ n \right] - {{\bf{q}}_{s,k}}} \right\|}^2}}$ obtained at the $r$th iteration, we have
\begin{align} \label{P2-31const3lowbound}
{\beta _{1,k}}\left[ n \right] \ge &\frac{{{\beta _0}}}{{{{\left( {{{\left\| {{\bf{q}}_u^r\left[ n \right] - {{\bf{q}}_{s,k}}} \right\|}^2} + {{\left( {{H_u} - {H_s}} \right)}^2}} \right)}^{{\alpha _1}/2}}}} - \frac{{{\alpha _1}{\beta _0}}}{{2{{\left( {{{\left\| {{\bf{q}}_u^r\left[ n \right] - {{\bf{q}}_{s,k}}} \right\|}^2} + {{\left( {{H_u} - {H_s}} \right)}^2}} \right)}^{\frac{{{\alpha _1}}}{2} + 1}}}}\notag\\
& \times \left( {{{\left\| {{{\bf{q}}_u}\left[ n \right] - {{\bf{q}}_{s,k}}} \right\|}^2}{\rm{ - }}{{\left\| {{\bf{q}}_u^r\left[ n \right] - {{\bf{q}}_{s,k}}} \right\|}^2}} \right)\overset{\triangle}{ =} {\varphi ^{lb}}\left( {{\beta _{1,k}}\left[ n \right]} \right).
\end{align}
Define the new constraint as
\begin{align}
{\varphi ^{lb}}\left( {{\beta _{1,k}}\left[ n \right]} \right) \ge {z_{1,k}}\left[ n \right],\forall k,n, \label{wsrP2-31const3lowboundadditional}
\end{align}
which is convex since ${\varphi ^{lb}}\left( {{\beta _{1,k}}\left[ n \right]} \right)$ is a  quadratic function  w.r.t. ${{{\bf{q}}_u}\left[ n \right]}$.
Similarly, for any local point ${{{\left\| {{\bf{q}}_u^r\left[ n \right] - {{\bf{q}}_{b}}} \right\|}^2}}$ obtained at the $r$th iteration, with ${\beta _3}\left[ n \right] = {{{\beta _0}} \over {d_3^{{\alpha _3}}\left[ n \right]}}$ and  

\noindent${d_3}\left[ n \right] = \sqrt {{{\left\| {{{\bf{q}}_u}\left[ n \right] - {{\bf{q}}_b}} \right\|}^2} + {{\left( {{H_b} - {H_u}} \right)}^2}} $,
$\beta_3[n]$ which is defined in  \eqref{wsrP2-31const4} can be replaced by
\begin{align}
&\frac{{{\beta _0}}}{{{{\left( {{{\left\| {{\bf{q}}_u^r\left[ n \right]{\rm{ - }}{{\bf{q}}_b}} \right\|}^2} + {{\left( {{H_u} - {H_b}} \right)}^2}} \right)}^{\frac{{{\alpha _3}}}{2}}}}} - \frac{{{\alpha _3}{\beta _0}}}{{2{{\left( {{{\left\| {{\bf{q}}_u^r\left[ n \right] - {{\bf{q}}_b}} \right\|}^2} + {{\left( {{H_u} - {H_b}} \right)}^2}} \right)}^{\frac{{{\alpha _3}}}{2} + 1}}}}\notag\\
&\qquad\qquad\qquad\qquad\qquad \times \left( {{{\left\| {{{\bf{q}}_u}\left[ n \right] - {{\bf{q}}_b}} \right\|}^2}{\rm{ - }}{{\left\| {{\bf{q}}_u^r\left[ n \right] - {{\bf{q}}_b}} \right\|}^2}} \right) \ge {z_3}\left[ n \right],\forall n,\label{wsrP2-31const4lowbound}
\end{align}
which is also a convex constraint.

In addition, to tackle the non-convexity of constraint \eqref{wsrP2-31const2}, we introduce variable $z_{2,k}[n]$, and reformulate \eqref{wsrP2-31const2} as
\begin{align}
&\sum\limits_{k = 1}^K {{a_k}\left[ n \right]} \left( {\rho {{\log }_2}\left( {1 + \frac{{P\left( {\left( {{{\rm{c}}_{k,1}}{\rm{ + }}{{\rm{c}}_{k,3}}} \right){z _{1,k}}\left[ n \right]{\rm{ + }}{{\rm{c}}_{k,2}}z_{2,k}[n] {\rm{ + }}{z _3}\left[ n \right]} \right)}}{{{\sigma ^2}}}} \right)} \right.\notag\\
&\qquad\qquad\qquad\qquad\qquad+{\left( {1 - \rho } \right){{\log }_2}\left( {1 + \frac{{P{z_3}\left[ n \right]}}{{{\sigma ^2}}}} \right)} \Bigg) \ge {R_{{\rm{th}}}},\forall n,\label{wsrP2-31const2NEW}
\end{align}
with the additional constraint
\begin{align}
{z_{1,k}}\left[ n \right] \ge \frac{{z_{2,k}^2\left[ n \right]}}{{{z_3}\left[ n \right]}},\forall k,n. \label{wsrP2-31const2NEWadditonal}
\end{align}
Both  constraints  \eqref{wsrP2-31const2NEW} and \eqref{wsrP2-31const2NEWadditonal} are convex since we can see that  the left hand side of \eqref{wsrP2-31const2NEW} is a log function, which is concave, and    the right hand side of \eqref{wsrP2-31const2NEWadditonal} is a  quadratic-over-linear fractional function, which is convex.  As a result, for any given local points  ${{{\left\| {{{\bf{q}}^r_u}\left[ n \right] - {{\bf{q}}_{s,k}}} \right\|}^2}}$ and ${{{\left\| {{{\bf{q}}^r_u}\left[ n \right] - {{\bf{q}}_{b}}} \right\|}^2}}$,  we have the following optimization problem

\begin{subequations}  \label{wsrP3}
	\begin{align}
	&\mathop {\max }\limits_{{{\bf{q}}_u}\left[ n \right],{z_{1,k}}\left[ n \right],{z_{2,k}}\left[ n \right],{z_3}\left[ n \right]} \sum\limits_{k = 1}^K {{w_k}\sum\limits_{n = 1}^N {{a_k}\left[ n \right]{\cal F}\left( {\frac{{\left( {{{\rm{c}}_{k,1}}{\rm{ + }}{{\rm{c}}_{k,3}}} \right){z_{1,k}}\left[ n \right]}}{{{\sigma ^2}}}} \right)} } \\
	&{\rm s.t.}~\eqref{Pconst5}, \eqref{Pconst6}, \eqref{wsrP2-31const3lowboundadditional}, \eqref{wsrP2-31const4lowbound}, \eqref{wsrP2-31const2NEW}, \eqref{wsrP2-31const2NEWadditonal}.
	\end{align}
\end{subequations}

\subsection{Convergence Analysis and Computational Complexity}
\begin{algorithm}[!t]
	\caption{Proposed relaxation-based algorithm for solving problem \eqref{wsrP1}.}
	\label{alg2}
	\begin{algorithmic}[1]
		\STATE  \textbf{Initialize} ${{{\left\| {{{\bf{q}}^r_u}\left[ n \right] - {{\bf{q}}_{s,k}}} \right\|}^2}}$,  $r=0$, $r_{\rm max}$.
		\STATE  Relax binary scheduling variables as continuous variables, and set $a_k^r[n]{\rm{ = }}{1 \mathord{\left/
				{\vphantom {1 K}} \right.
				\kern-\nulldelimiterspace} K}$.
		\STATE  \textbf{Repeat}
		\STATE  \qquad Solve problem \eqref{wsrP1-1}  for given $\{{\bf q}^r_u[n]\}$, and denote the optimal solution as $\{a^{r+1}_k[n]\}$.
		\STATE  \qquad Solve problem \eqref{wsrP3}  for given $\{a^{r+1}_k[n]\}$, and denote the optimal solution as $\{{\bf q}^{r+1}_u[n]\}$.
		\STATE  \qquad  Update $r\leftarrow r+1$.
		\STATE \textbf{Until} the fractional increase in the objective value of \eqref{wsrP1} is below a threshold or the \\
		maximum number of  iterations $r_{\rm max}$ is reached.
	\end{algorithmic}
\end{algorithm}
In the proposed AO algorithm, we solve the relaxed  problem  \eqref{wsrP1} by iteratively solving problems \eqref{wsrP1-1} and \eqref{wsrP3}, where the solution obtained for one subproblem in each iteration is used as the initial point for the other. The detailed procedure for solving \eqref{wsrP1} is summarized in Algorithm~\ref{alg2}.  The convergence of Algorithm~\ref{alg2} has been well studied in \cite{Bertsekas1997Nonlinear}, and is omitted here for brevity.

We now analyze the complexity of Algorithm~\ref{alg2}. In step 4, \eqref{wsrP1-1} is a linear optimization problem whose complexity is ${\cal O}\left( {KN} \right)$ \cite{gondzio1996computational}, where $KN$ denotes the number of variables. In step 5, the complexity  for solving \eqref{wsrP3}  by the interior point method is ${\cal O}{\left( {2KN+3N} \right)^{3.5}}$ \cite{zhang2019securing}, where ${ 2KN+3N}$ denotes the number of variables. Therefore, the total complexity of Algorithm~\ref{alg2}  is ${\cal O}\left( {{L_{{\rm{iter}}}}\left( {KN{\rm{ + }}{{\left( {2KN + 3N} \right)}^{3.5}}} \right)} \right)$, where  ${{L_{{\rm{iter}}}}}$  stands for  the  number of iterations required to reach convergence.
\section{Penalty-based Algorithm for Fairness BER Optimization  Problem}
In this section, we aim to solve  problem~\eqref{P}. Based on Theorem 2, \eqref{SectionIII_3}, and \eqref{SectionIII_4} in Section III, problem \eqref{P} is simplified as
\begin{subequations} \label{P1}
	\begin{align}
	&\mathop {\max }\limits_{{{\bf{q}}_u}\left[ n \right],{a_k}\left[ n \right],R} R \label{SectionIII_5}\\
	&{\rm{s}}.{\rm{t}}.\frac{1}{N}\sum\limits_{n = 1}^N {{a_k}\left[ n \right]{\cal F}\left( {\frac{{\left( {{{\rm{c}}_{k,1}}{\rm{ + }}{{\rm{c}}_{k,3}}} \right){\beta _{1,k}}\left[ n \right]}}{{{\sigma ^2}}}} \right) \ge R,\forall k}, \label{max-minP1const1}\\
	&\qquad\sum\limits_{k = 1}^K {{a_k}\left[ n \right]} \left( {\rho {{\log }_2}\left( {1 + \frac{{P\left( {\left( {{{\rm{c}}_{k,1}}{\rm{ + }}{{\rm{c}}_{k,3}}} \right){\beta _{1,k}}\left[ n \right]{\rm{ + }}{{\rm{c}}_{k,2}}\sqrt {{\beta _{1,k}}\left[ n \right]{\beta _3}\left[ n \right]} {\rm{ + }}{\beta _3}\left[ n \right]} \right)}}{{{\sigma ^2}}}} \right)} \right.\notag\\
	&\qquad +{\left( {1 - \rho } \right){{\log }_2}\left( {1 + \frac{{P{\beta _3}\left[ n \right]}}{{{\sigma ^2}}}} \right)} \Bigg) \ge {R_{{\rm{th}}}},\forall n,\label{SectionIII_6}\\
	&\qquad  \eqref{Pconst5}, \eqref{Pconst6}, \eqref{Pconst2}, \eqref{Pconst3}.
	\end{align}
\end{subequations}
Unfortunately, the low-complexity algorithm based on the relaxation-based  method cannot be applied to  problem~\eqref{P1} due to the primary rate requirement~\eqref{SectionIII_6}. More specifically, when converting the continuous-valued solutions for the $a_k[n]$ obtained by the relaxed problem to binary, e.g., using  the rounding function \cite{HUA2020throughput}, constraint~\eqref{SectionIII_6} will in general no longer be satisfied. In this section, we propose a two-layer penalty-based algorithm to solve \eqref{P1}.  The
inner layer solves a penalized optimization problem by applying the AO
method, while the outer layer updates the penalty coefficient, until convergence is achieved. Specifically, in the inner layer, the original problem \eqref{P1} is  decomposed into three subproblems: IRS phase shift matrix optimization, IRS scheduling optimization, and UAV trajectory optimization. 

We first introduce slack variables $\{{\bar{a}}_k[n]\}$ to transform the binary constraints into a series of equivalent equality constraints. Specifically, \eqref{Pconst3} can be rewritten as
\begin{align}
{a_k}\left[ n \right]\left( {1{\rm{ - }}{{\bar a}_k}\left[ n \right]} \right){\rm{ = }}0,\forall k,n, \label{SectionIII_7}
\end{align}
\begin{align}
{\kern 1pt} {a_k}\left[ n \right] = {{\bar a}_k}\left[ n \right],\forall k,n. \label{SectionIII_8}
\end{align}
From \eqref{SectionIII_7} and \eqref{SectionIII_8}, we can readily derive that  the $a_k[n]$ that satisfies the above two constraints must be either $1$ or $0$, which confirms the equivalence of the transformation of \eqref{Pconst3} into the two constraints. We then use \eqref{SectionIII_7} and  \eqref{SectionIII_8} in a penalty term that is added to the objective function of \eqref{P1}, yielding the following optimization problem
\begin{subequations} \label{P2}
\begin{align}
&\mathop {\min }\limits_{{{\bf{q}}_u}\left[ n \right],{a_k}\left[ n \right],R,{{\bar a}_k}\left[ n \right]}  - R{\rm{ + }}\frac{1}{{2\eta }}\sum\limits_{k = 1}^K {\sum\limits_{n = 1}^N {\left( {{{\left| {{a_k}\left[ n \right]\left( {1{\rm{ - }}{{\bar a}_k}\left[ n \right]} \right)} \right|}^2}{\rm{ + }}{{\left| {{a_k}\left[ n \right]{\rm{ - }}{{\bar a}_k}\left[ n \right]} \right|}^2}} \right)} } \\
&\qquad{\rm s.t.}~ \eqref{Pconst5}, \eqref{Pconst6},\eqref{Pconst2} ,\eqref{max-minP1const1},\eqref{SectionIII_6} 
\end{align}
\end{subequations}
where $\eta>0 $ is the penalty coefficient used to penalize the violation of the equality constraints \eqref{SectionIII_7} and \eqref{SectionIII_8} \cite{Bertsekas1997Nonlinear}. While these equality constraints become satisfied as $\eta \to 0$, it is not effective to initially set $\eta$ to be a very small value since in this case the objective will be dominated by the penalty terms, and the   term $-R$ will be diminished. In contrast, initializing $\eta$ with a larger value allows us to obtain a good starting point for the proposed algorithm.  Then, by gradually decreasing the value of $\eta$, we can finally obtain a solution that satisfies \eqref{SectionIII_7} and \eqref{SectionIII_8}  within a predefined accuracy. Note that, for any given penalty coefficient $\eta$, problem \eqref{P2} is still non-convex due to the non-convex  constraints \eqref{max-minP1const1} and \eqref{SectionIII_6}. We then   apply the AO method to iteratively optimize the primary variables in different blocks \cite{Bertsekas1997Nonlinear}. Specifically, in the inner layer, problem \eqref{P2}
is divided into three subproblems in which $\{\bar a_k[n]\}$,  $\{a_k[n]\}$, and $\{{\bf q}_u[n]\}$ are optimized iteratively as follows:
\subsection{Inner layer iteration}
1) Optimizing $\bar a_k[n]$ for given $a_k[n]$ and ${\bf q}_u[n]$.  This subproblem can be expressed as
\begin{subequations} \label{P2-1}
\begin{align}
&\mathop {\min }\limits_{{R},{{\bar a}_k}\left[ n \right]}  - R{\rm{ + }}\frac{1}{{2\eta }}\sum\limits_{k = 1}^K {\sum\limits_{n = 1}^N {\left( {{{\left| {{a_k}\left[ n \right]\left( {1{\rm{ - }}{{\bar a}_k}\left[ n \right]} \right)} \right|}^2}{\rm{ + }}{{\left| {{a_k}\left[ n \right]{\rm{ - }}{{\bar a}_k}\left[ n \right]} \right|}^2}} \right)} }\\
&\qquad{\rm s.t.}~\eqref{max-minP1const1}.
\end{align}
\end{subequations}
We can see that only the auxiliary variable $\bar a_k[n]$ is involved in the objective function. Therefore, setting the derivative of \eqref{P2-1}  w.r.t. $\bar a_k[n]$ to zero, the solution can be  obtained as
\begin{align}
\bar a_k^{{\rm{opt}}}\left[ n \right] = \frac{{{a_k}\left[ n \right] + a_k^2\left[ n \right]}}{{1 + a_k^2\left[ n \right]}},\forall k,n. \label{P2-2bara}
\end{align}
\hspace*{\parindent}2) Optimizing $ a_k[n]$ for given $\bar a_k[n]$ and ${\bf q}_u[n]$.  This subproblem is written as
\begin{subequations} \label{P2-2}
	\begin{align}
	&\mathop {\min }\limits_{{{ a}_k}\left[ n \right],R}  - R{\rm{ + }}\frac{1}{{2\eta }}\sum\limits_{k = 1}^K {\sum\limits_{n = 1}^N {\left( {{{\left| {{a_k}\left[ n \right]\left( {1{\rm{ - }}{{\bar a}_k}\left[ n \right]} \right)} \right|}^2}{\rm{ + }}{{\left| {{a_k}\left[ n \right]{\rm{ - }}{{\bar a}_k}\left[ n \right]} \right|}^2}} \right)} }\\
	& \qquad{\rm s.t.}~\eqref{Pconst2},\eqref{max-minP1const1},\eqref{SectionIII_6}.
	\end{align}
\end{subequations}
It can be seen that \eqref{P2-2} is convex with  a quadratic objective function and linear inequality constraints, which can be numerically solved by standard convex optimization
techniques, such as the interior-point method \cite{boyd2004convex}.

3) Optimizing ${\bf q}_u[n]$ for given $a_k[n]$ and $\bar a_k[n]$. Ignoring the constant terms that are irrelevant to the UAV trajectory,  this subproblem is formulated as:
\begin{subequations} \label{P2-3}
	\begin{align}
	&\mathop {\max }\limits_{{{\bf{q}}_u}\left[ n \right],R} R\\
	&\qquad{\rm s.t.}~\eqref{Pconst5}, \eqref{Pconst6},\eqref{max-minP1const1},\eqref{SectionIII_6}.
	\end{align}
\end{subequations}
Note that \eqref{P2-3} is neither concave or quasi-concave due to the non-convex   constraints \eqref{max-minP1const1} and  \eqref{SectionIII_6}. In general, there is no efficient method
to obtain the optimal solution. In the following, we adopt the
successive convex optimization technique to solve \eqref{P2-3}. 
Using the previous analysis of the UAV trajectory  optimization for  problem \eqref{wsrP2} in Section III-B, by introducing the same slack variables $\{z_{1,k}[n],z_{2,k}[n],z_{3}[n]\}$ and local points ${{{\left\| {{{\bf{q}}^r_u}\left[ n \right] - {{\bf{q}}_{s,k}}} \right\|}^2}}$ and ${{{\left\| {{{\bf{q}}^r_u}\left[ n \right] - {{\bf{q}}_{b}}} \right\|}^2}}$, we can directly derive the  following equivalent convex optimization problem

\begin{subequations} \label{P2-32}
	\begin{align}
	&\mathop {\max }\limits_{{{\bf{q}}_u}\left[ n \right],{z_{1,k}}\left[ n \right],{z_{2,k}}\left[ n \right], {z_3}\left[ n \right],R} R \\
	&{\rm{s}}.{\rm{t}}.\;\frac{1}{N}\sum\limits_{n = 1}^N {{a_k}\left[ n \right]{\cal F}\left( {\frac{{\left( {{{\rm{c}}_{k,1}}{\rm{ + }}{{\rm{c}}_{k,3}}} \right){z_{1,k}}\left[ n \right]}}{{{\sigma ^2}}}} \right) \ge R,\forall k},\\
	&\qquad \eqref{Pconst5}, \eqref{Pconst6},\eqref{wsrP2-31const3lowboundadditional}, \eqref{wsrP2-31const4lowbound}, \eqref{wsrP2-31const2NEW}, \eqref{wsrP2-31const2NEWadditonal}.
	\end{align}
\end{subequations}
Based on the previous discussions,  the objective function and all of the constraints are convex. Thus, \eqref{P2-32} is a convex optimization problem that can be efficiently solved by, for example, the interior point  method \cite{boyd2004convex}.

\subsection{Outer layer iteration}
In the outer layer, we gradually decrease the value of the penalty coefficient $\eta$ as follow
\begin{align}
\eta=c\eta, \label{penaltcoefficentupdate}
\end{align}
where $c$ $(0<c<1)$ is a  scaling factor, where   a larger value of $c$ can achieve
better performance but at the cost of more iterations in the outer layer.

\subsection{Convergence Analysis and Computational Complexity}
To show the converged solutions of the proposed penalty-based algorithm, the terminal criteria for the outer layer is given as follows;
\begin{align}
\xi {\rm{ = }}\max \left\{ {\left| {{a_k}\left[ n \right]\left( {1{\rm{ - }}{{\bar a}_k}\left[ n \right]} \right)} \right|,\left| {{a_k}\left[ n \right]{\rm{ - }}{{\bar a}_k}\left[ n \right]} \right|,\forall k,n} \right\}, \label{constraintviolation}
\end{align}
where $\xi$ is a  predefined accuracy. The detailed procedure of the penalty-based algorithm is summarized in Algorithm~\ref{alg1}.
\begin{algorithm}[!t]
	\caption{Proposed penalty-based algorithm for solving problem \eqref{P1}.}
	\label{alg1}
	\begin{algorithmic}[1]
		\STATE  \textbf{Initialize} $a^{r_1}_k[n]$, ${{{\left\| {{{\bf{q}}^{r_1}_u}\left[ n \right] - {{\bf{q}}_{s,k}}} \right\|}^2}}$, $\eta$, $r_1=0$, $r_2=0$ $\varepsilon_1$, $\varepsilon_2$, $r_{\rm max}$.
		\STATE  \textbf{Repeat: outer layer}
		\STATE \quad \textbf{Repeat: inner layer }
		\STATE  \qquad Update $\bar a^{r_1}_k[n]$ based on \eqref{P2-2bara}.
		\STATE  \qquad Update $ a^{r_1}_k[n]$ by solving problem \eqref{P2-2}.
		\STATE  \qquad Update ${\bf q}^{r_1}_u[n]$ by solving problem \eqref{P2-32}.
		\STATE  \qquad $r_1\leftarrow r_1+1$.
      	\STATE \quad \textbf{Until} the fractional decrease of the objective value of \eqref{P2} is below a threshold $\varepsilon_1$ or the \\
      	\quad  maximum number of
      	iterations $r_{\rm max}$ is reached.
		\STATE  \quad Update penalty coefficient $\eta^{r_2}$ based on \eqref{penaltcoefficentupdate}.
		\STATE  \quad $r_2\leftarrow r_2+1$, and $r_1\leftarrow 0$.
		\STATE \textbf{Until} the constraint violation $\xi $ is below a threshold $\varepsilon_2$
%		\STATE {\bf Output:}  max-min rate $R$ and IRS shift phase ${\theta^{\rm opt} _{k,m}}\left[ n \right]$ according to \eqref{theorem2}.
	\end{algorithmic}
\end{algorithm}
In the inner layer, with the given penalty coefficient, the objective function of \eqref{P2} is non-increasing over each iteration after applying the AO method, and the objective of \eqref{P2} is bounded due to the limited flying time $T$ and transmit power $P$. As such, a  stationary point can be achieved in the inner layer. In the outer layer, we gradually decrease the penalty coefficient so that the equality constraints  \eqref{SectionIII_7} and \eqref{SectionIII_8} are ultimately satisfied.  Based on the results in \cite[Appendix B]{joint2017cai}, this penalty-based framework is guaranteed to converge.

The complexity of Algorithm~\ref{alg1} can be quantified as follows. In the inner layer, the main complexity of Algorithm~\ref{alg1} comes from steps 5 and 6. In step 5, the complexity of computing $a_k[n]$  is ${\cal O}{\left( { KN+2N +1} \right)^{3.5}}$ \cite{zhang2019securing}, where $KN+2N +1$ stands for the number of  variables \cite{boyd2016EE}. Similarly,  in step 6,  the complexity required to compute the UAV trajectory  is  ${\cal O}{\left( {2KN+3N + 1} \right)^{3.5}}$ \cite{zhang2019securing}, where ${ 2KN+3N +1}$ denotes the number of variables. Therefore, the total complexity of Algorithm~\ref{alg1}  is ${\cal O}\left( {{L_{{\rm{outer}}}}{L_{{\rm{inner}}}}\left( {{{\left( {KN + 2N + 1} \right)}^{3.5}}{\rm{ + }}{{\left( {2KN + 3N + 1} \right)}^{3.5}}} \right)} \right)$, where  ${{L_{{\rm{inner}}}}}$ and ${{L_{{\rm{outer}}}}}$ respectively  denote the  number of iterations required for reaching convergence in the inner layer and outer layer.

\section{Numerical Results}
In this section, we provide numerical results  to verify the performance of the proposed algorithm  for the UAV assisted IRS  symbiotic radio  transmission system.  In the simulation, we consider a system that operates on a carrier frequency of $755~{\rm MHz}$ with the system bandwidth of $1~{\rm MHz}$ and the effective noise power density   $-120 {\rm  dBm/Hz}$. As such,  the noise power at the BS and the channel gain are set to  ${\sigma ^2} =  - 60~{\rm{dBm}}$ and ${\beta _0} =  - 30~{\rm{dB}}$, respectively  \cite{xu2018uav}. In addition, we set  ${{\rm{d}} \mathord{\left/
		{\vphantom {{\rm{d}} \lambda }} \right.
		\kern-\nulldelimiterspace} \lambda } = {1 \mathord{\left/
		{\vphantom {1 2}} \right.
		\kern-\nulldelimiterspace} 2}$ \cite{joint2020wu}. 
 The UAV altitude is fixed at
$H_u=30~\rm m$ with transmit power $P=20~ \rm{dBm}$ and maximum speed  $V_{\max}=10~{\rm m/s}$. The UAV's initial and final location are set to ${{\bf{q}}_{\rm{I}}}={{\bf{q}}_{\rm{F}}}= {\left[ {15{\rm{m}}{\kern 1pt} {\kern 1pt} {\kern 1pt} {\kern 1pt} {\kern 1pt} 0} \right]^T}$. The altitudes of the BS and IRS are both set to  $H_s=H_b=10~\rm m$. The duration of each time slot is $\delta  = 0.1~\rm s$. The path loss exponents for the UAV-IRS link, IRS-BS link, and UAV-BS link are assumed to be the same $2.4$, and   the Rician factors for the above links are set to be  $10~{\rm dB}$. Without loss of generality, we set the utility function ${\cal F}(\cdot)$ as a logarithm function with base $2$,  which naturally achieves a certain of fairness among the information transmission of multiple IRSs, and has been widely adopted in the literature, such as \cite{ye2013User}.
Unless  otherwise  specified, we set $r_{\max}=300$, $\rho=0.5$,  ${\varepsilon _1} = {10^{ - 3}}$, ${\varepsilon _2} = {10^{ - 10}}$,  $\eta  = 500$, $c=0.7$.
\subsection{Weighted Sum BER Optimization }
\begin{figure}[htbp]
	\centering
	\begin{minipage}[t]{0.46\textwidth}
		\centering
		\includegraphics[width=3.2in]{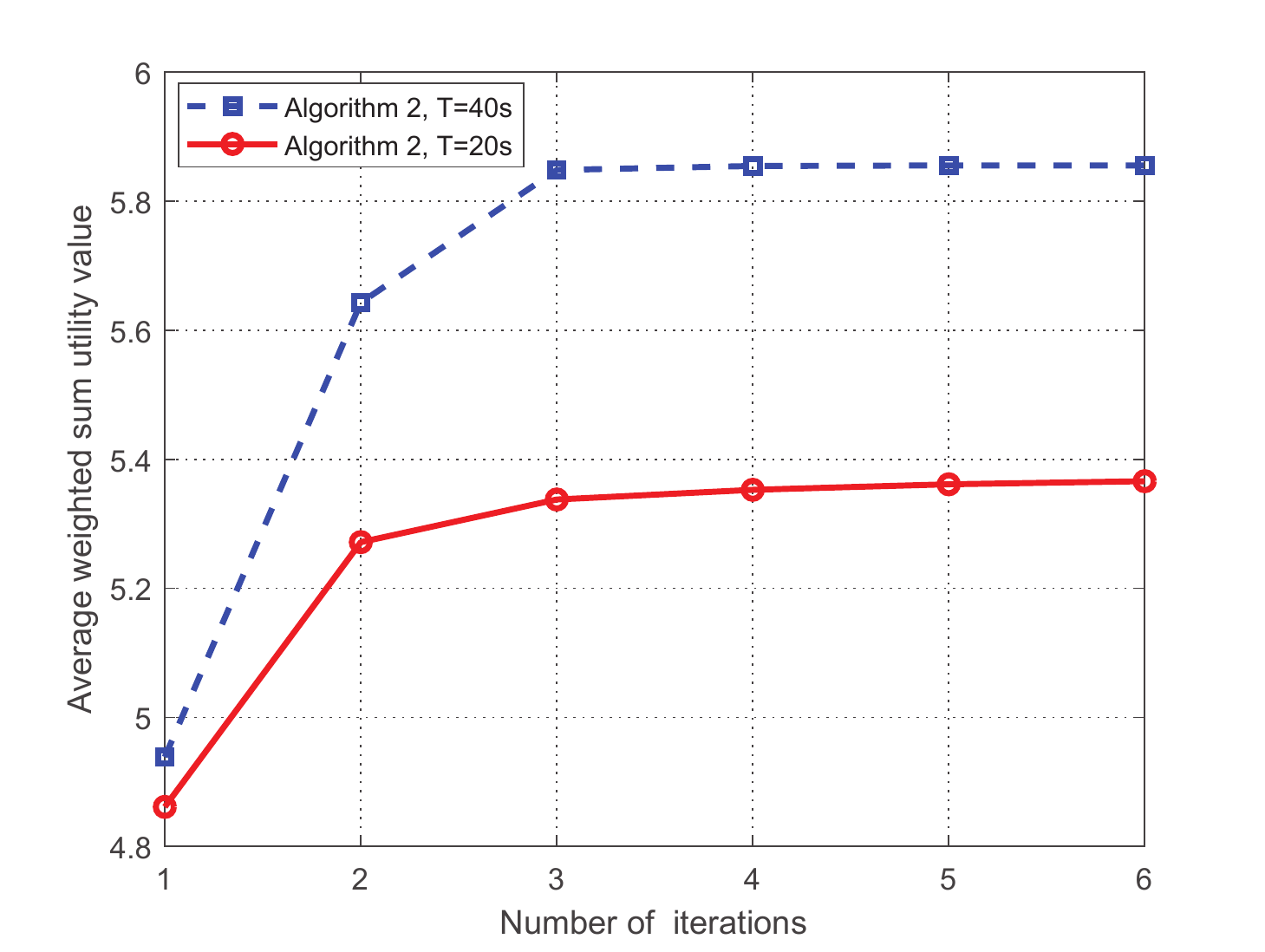}
		\caption{Convergence behaviour of the proposed Algorithm~\ref{alg2} for different period.}\label{WSRfig1}
	\end{minipage}
	\begin{minipage}[t]{0.46\textwidth}
		\centering
		\includegraphics[width=3.2in]{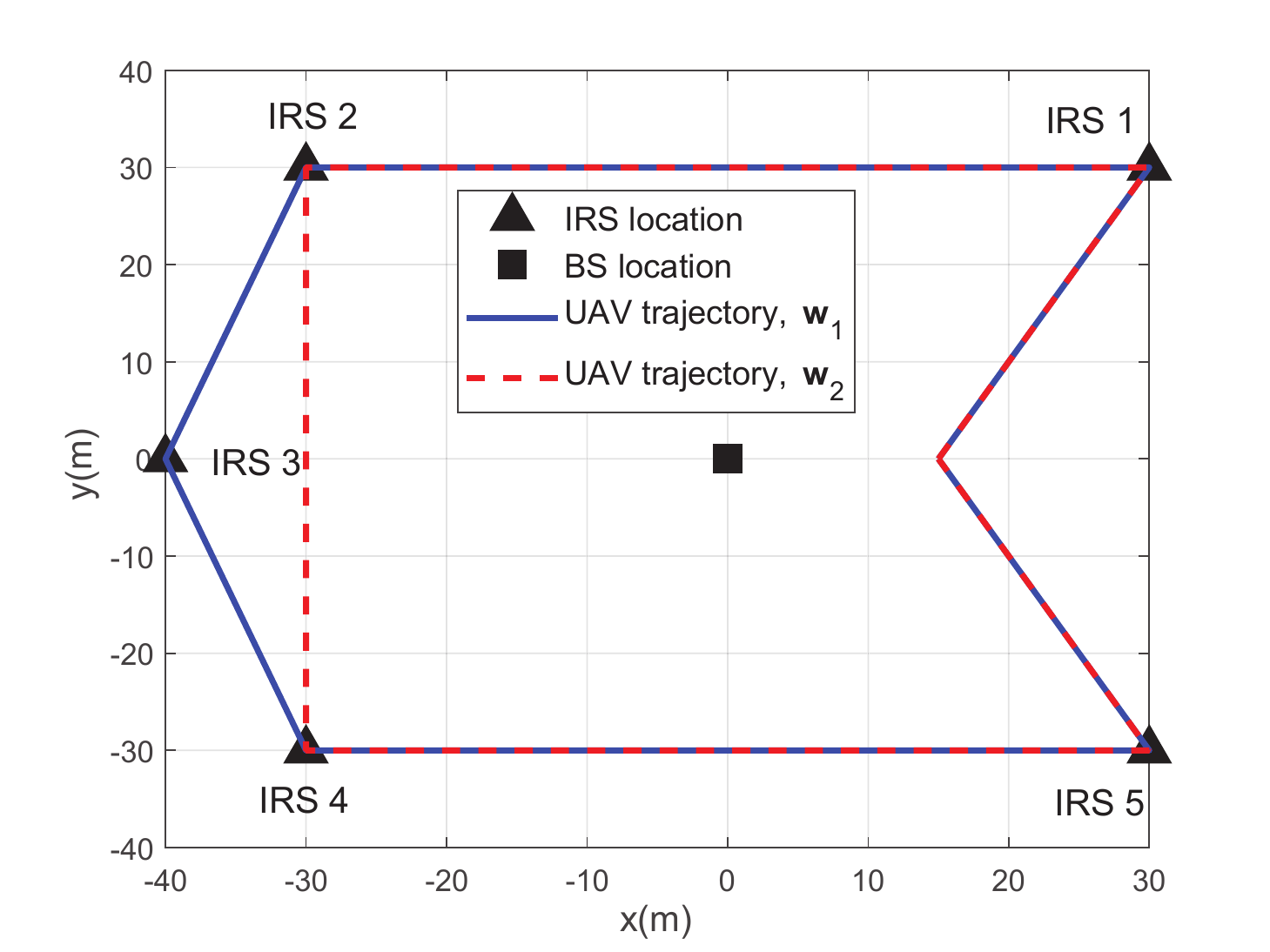}
		\caption{Optimized UAV trajectory for different weighting factors.} \label{WSRfig2}
	\end{minipage}
\end{figure}

 This subsection evaluates the performance of Algorithm~\ref{alg2} for the weighted sum BER problem \eqref{wsrP}. We consider $5$ IRS, which are located at ${{\bf{q}}_{s,1}} = {\left[ {30~{\rm{m}},30~{\rm{m}}} \right]^T},{{\bf{q}}_{s,2}} = {\left[ { - 30~{\rm{m}},30~{\rm{m}}} \right]^T},{{\bf{q}}_{s,3}} = {\left[ { - 40~{\rm{m}},0} \right]^T},{{\bf{q}}_{s,4}} = {\left[ { - 30~{\rm{m}}, - 30~{\rm{m}}} \right]^T},{{\bf{q}}_{s,5}} = {\left[ {30~{\rm{m}}, - 30~{\rm{m}}} \right]^T}$ in a horizontal plane.  Unless otherwise specified, the weighting factors are set as ${\bf{w}} = {\left[ {1,1,1,1,1} \right]^T}$. To show the efficiency  of  Algorithm~\ref{alg2},  its convergence behaviour for the two different periods $T$ is plotted in Fig.~\ref{WSRfig1}. It is observed that the average weighted sum utility value  increases quickly with the number of iterations, and in both cases converges within only $3$ iterations.

\begin{figure}[htbp]
	\centering
	\begin{minipage}[t]{0.48\textwidth}
		\centering
		\includegraphics[width=3.2in]{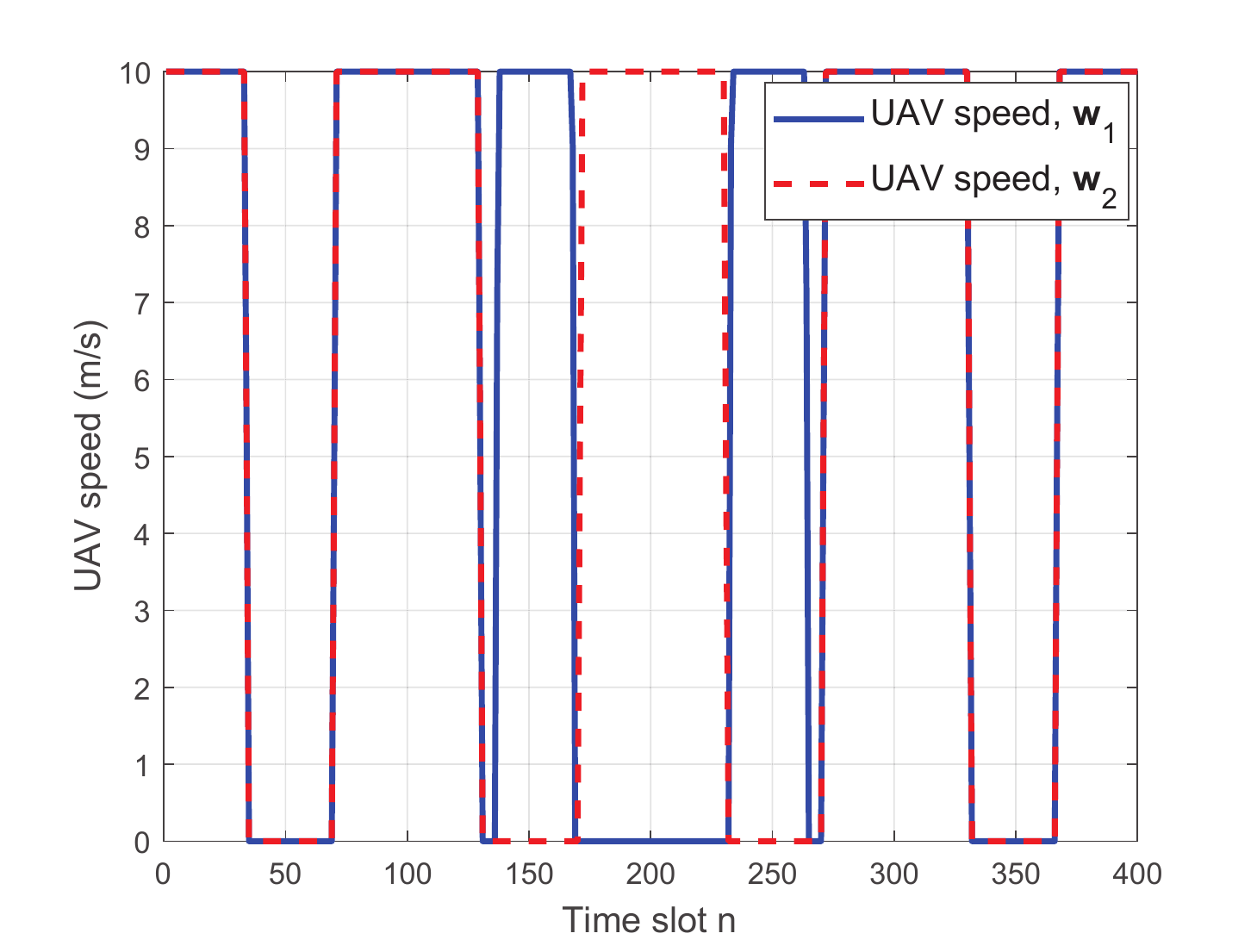}
		\caption{UAV speed  under  $T=40~\rm s$.}\label{WSRfig3}
	\end{minipage}
	\begin{minipage}[t]{0.48\textwidth}
		\centering
		\includegraphics[width=3.2in]{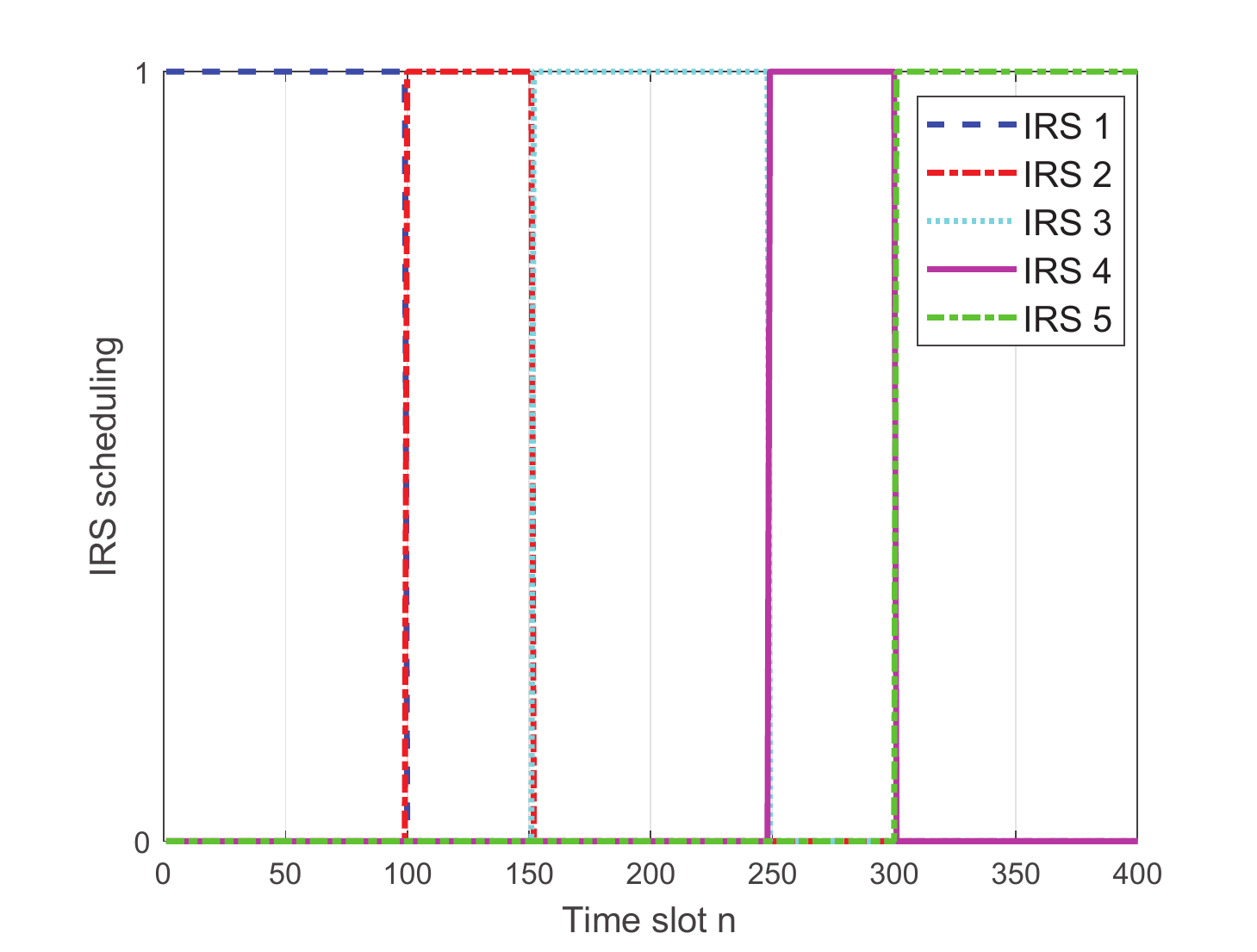}
		\caption{IRS scheduling under  $T=40~\rm s$.} \label{WSRfig4}
	\end{minipage}
\end{figure}

In Fig.~\ref{WSRfig2}, the optimized UAV trajectories obtained by Algorithm~\ref{alg2} when $T=40~\rm s$ are studied for two different weighting factors, i.e., ${{\bf{w}}_1} = {\left[ {1,1,1,1,1} \right]^T}$ and ${{\bf{w}}_2} = {\left[ {1,1,0.5,1,1} \right]^T}$. We see that the UAV sequentially visits all IRS for the weighting factor ${\bf{w}}_1$, since the path loss between the UAV and IRS is significantly reduced when the UAV is nearby, thereby improving the utility value. However, for weighting factor ${{\bf w}_2}$, the UAV only does a close fly-by of IRS~3 rather than hovering above it, since ${{\bf w}_2}$ places a lower weight on IRS~3 and hence reduces its priority relative to the others. To see this more clearly,  in Fig.~\ref{WSRfig3} the UAV speed profile for the two weighting factors is plotted. Compared  with ${{\bf w}_1}$, for ${{\bf w}_2}$ the UAV spends less time hovering above IRS $3$ for serving.

In Fig.~\ref{WSRfig4}, the IRS scheduling for $T=40~\rm s$ is plotted. We see that for optimizing the weighted sum utility, the IRSs are scheduled for different lengths of time as shown in Fig.~\ref{WSRfig4}. As before, the IRS scheduling results are indeed binary, which verifies the effectiveness of Algorithm~\ref{alg2}.
\begin{figure}[htbp]
	\centering
	\begin{minipage}[t]{0.45\textwidth}
		\centering
		\includegraphics[width=3.2in]{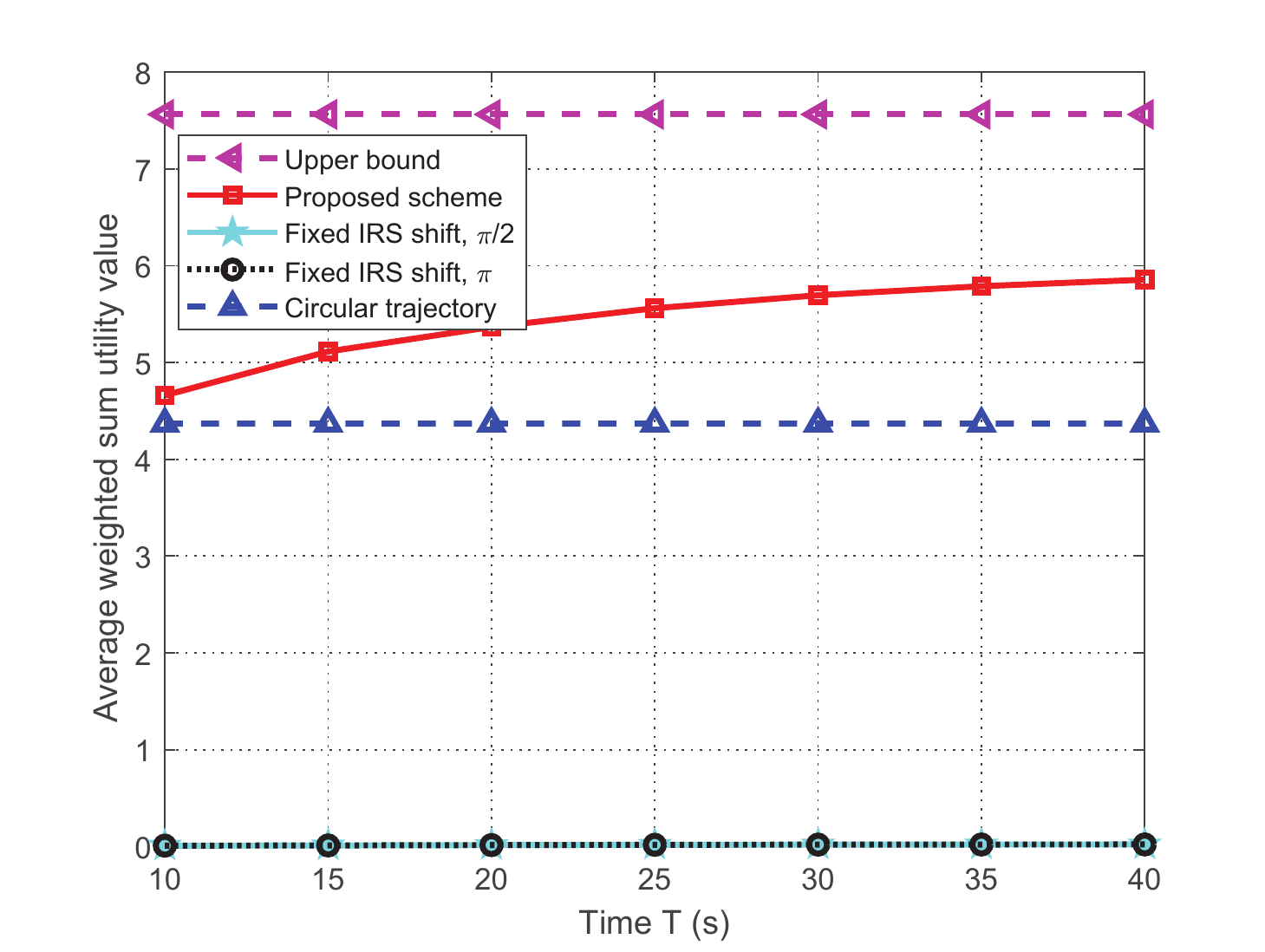}
		\caption{Average weighted sum utility value  versus period $T$.}\label{WSRfig5}
	\end{minipage}
	\begin{minipage}[t]{0.45\textwidth}
		\centering
		\includegraphics[width=3.2in]{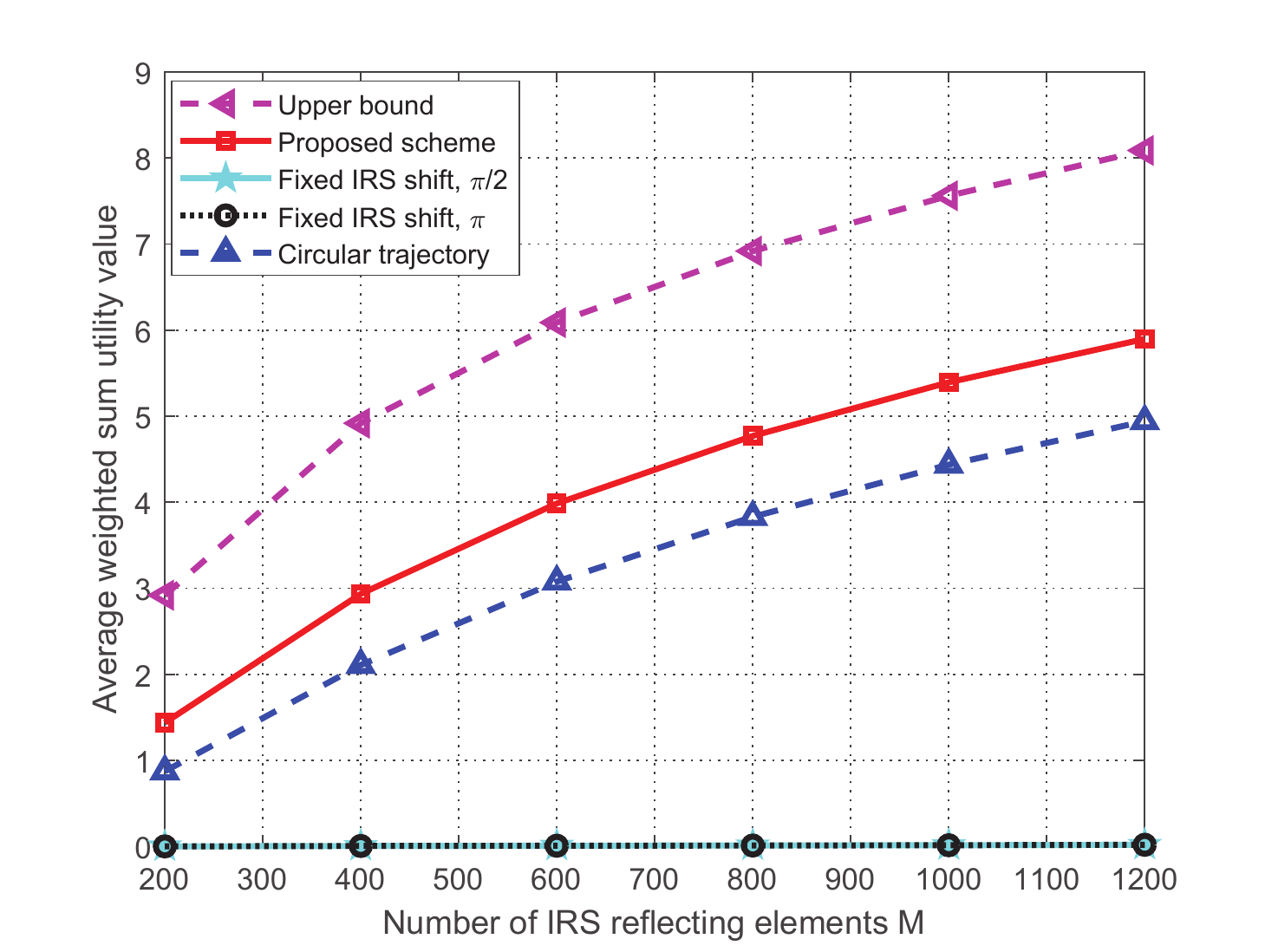}
		\caption{Average weighted sum utility value versus   the number of IRS reflecting elements.} \label{WSRfig6}
	\end{minipage}
\end{figure}

In Fig.~\ref{WSRfig5}, we  compares the average weighted sum utility value versus $T$ achieved by the following schemes: 1) Proposed scheme in Algorithm~\ref{alg2};  2) Circular trajectory, where the UAV flies with a circle path of radius $15~\rm m$ (corresponding to the distance from the BS to the UAV's initial/final location) and center ${\left[ {0, - 0} \right]^T}$; 3) Fixed  phase shifts, where the IRS phase shifts for all the elements is fixed at either $\pi$ or $\pi/2$. For the  fixed phase shift examples, the UAV trajectory is set to be the result obtained by the proposed scheme. 
 The upper bound for the weighted sum BER problem  is given by the solution to 
\begin{equation}
\mathop {\max }\limits_{\forall k} \left\{ {{{\log }_2}\left( {1 + \frac{{LP\left( {{c_{k,1}} + {c_{k,3}}} \right){\beta _0}}}{{{\sigma ^2}{{\left( {{H_u} - {H_s}} \right)}^{{\alpha _1}}}}}} \right)} \right\}.
\end{equation}
It is  observed from Fig.~\ref{WSRfig5} that our proposed algorithm substantially outperforms the other methods in terms of average weighted sum utility value. This is expected since an optimized UAV trajectory can establish better channel conditions for the IRS, which significantly increases IRS's SNR. In addition, by adjusting the IRS phase shifts to align the cascaded AoA and AoD with the UAV-BS link, i.e., as shown in Theorem~2,  the SNR of the UAV-IRS-BS link will be significantly increased.

In Fig.~\ref{WSRfig6},  we  study the average weighted sum utility value versus the number of IRS reflecting elements $M$. The performance gain of the proposed approach and the circular trajectory increases with $M$, since  more  reflecting elements help achieve higher passive beamforming gain. In addition, our proposed approach outperforms the circular trajectory by  leveraging the UAV mobility. Clearly, the IRS has a significant impact on the system performance, and the IRS phase shifts must be finely tuned in the system design.

\subsection{Fairness BER  Optimization}
\begin{figure*}[htbp]
	\centering
	\subfigure[Constraint violation $\xi$.]{
		\begin{minipage}[t]{0.45\linewidth}
			\centering
			\includegraphics[width=3.2in]{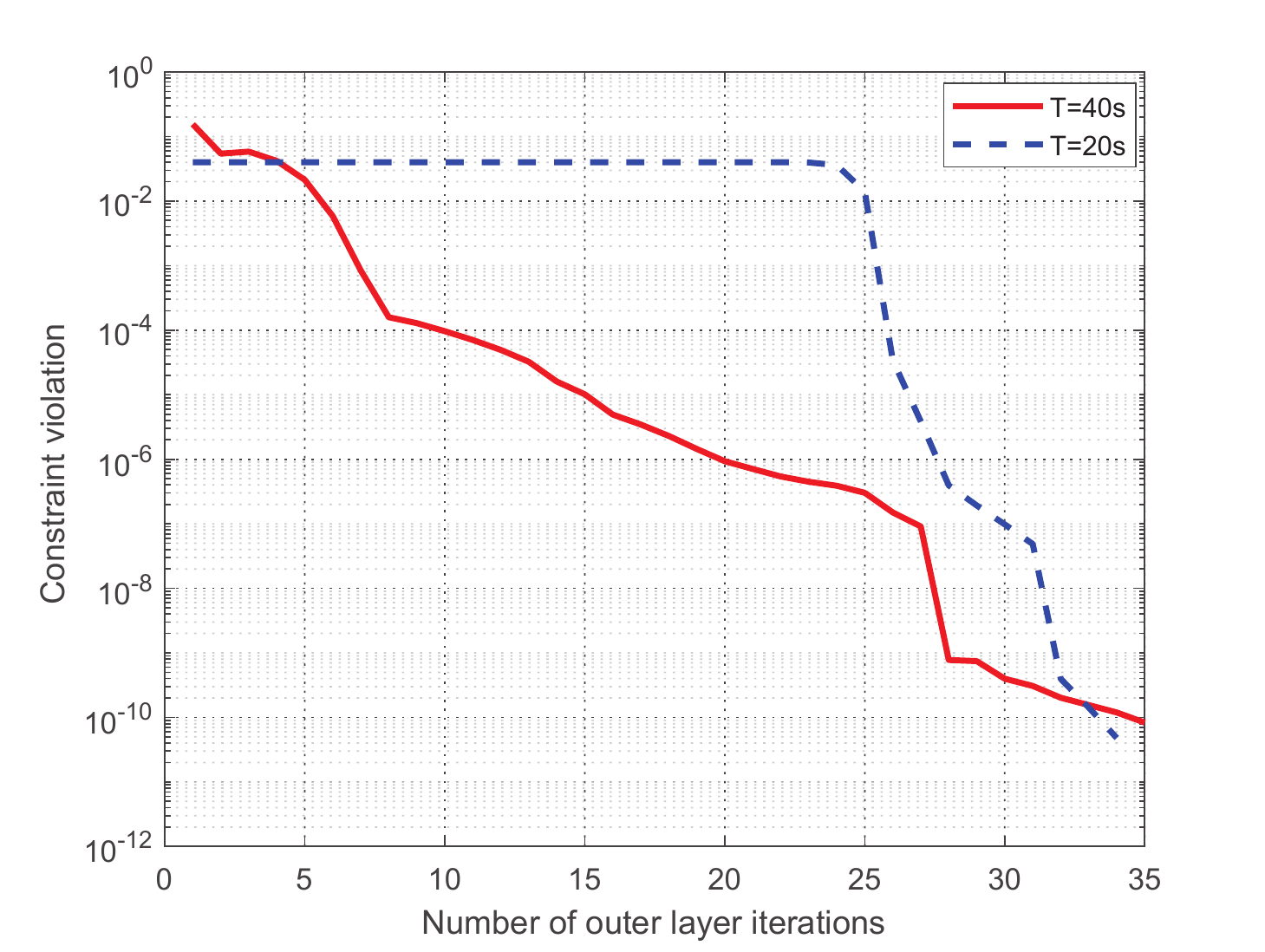}
			%\caption{fig1}
		\end{minipage}%
	}%
	\subfigure[Fairness utility value.]{
		\begin{minipage}[t]{0.45\linewidth}
			\centering
			\includegraphics[width=3.2in]{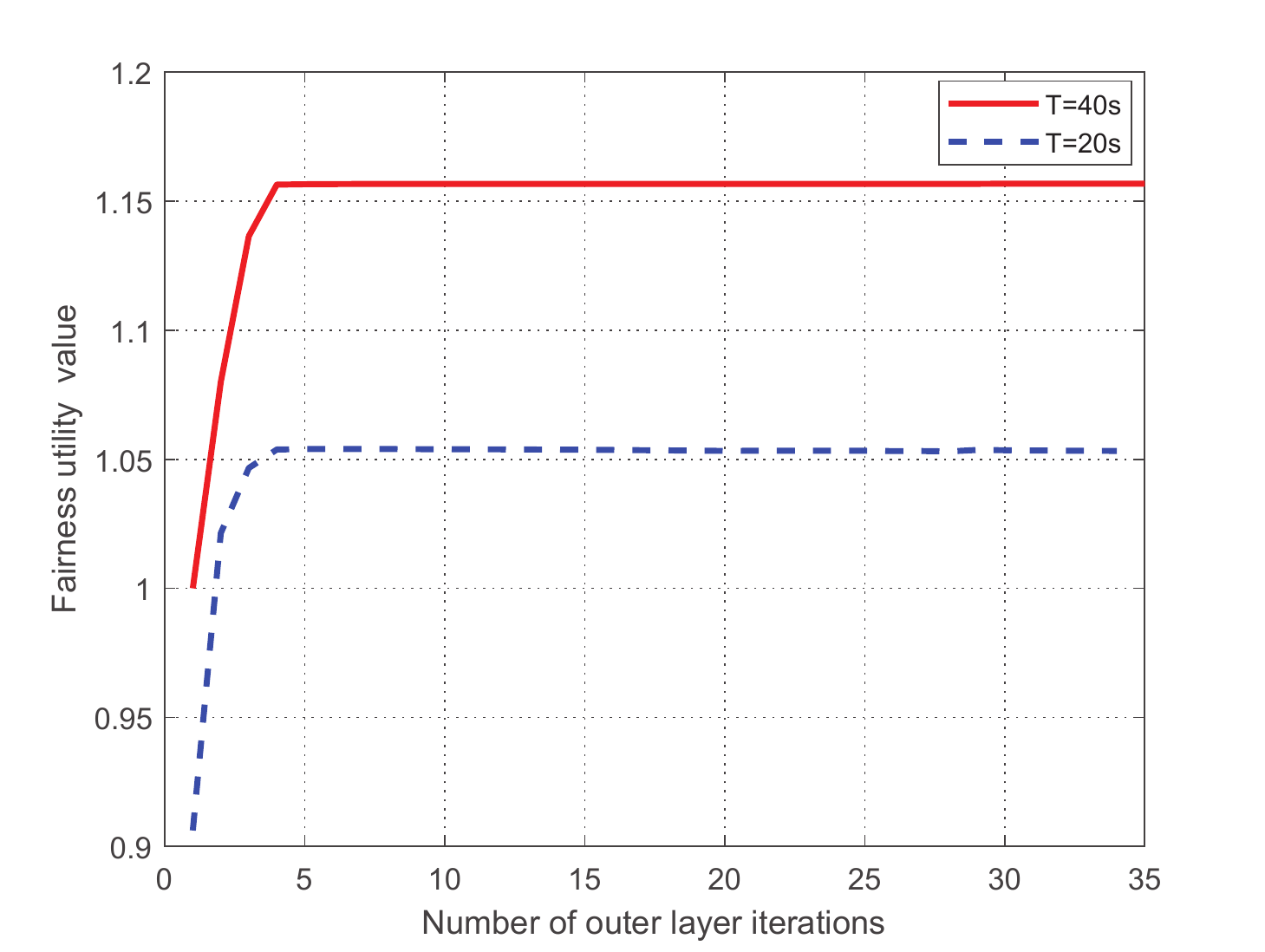}
			%\caption{fig2}
		\end{minipage}%
	}%
	\quad               %Õâ¸ö»Ø³µ¼üºÜÖØÒª \quadÒ²¿ÉÒÔ
	\centering
	\caption{Convergence behaviour of Algorithm~\ref{alg1}.}\label{maxminfig1}
\end{figure*}
This subsection evaluates the performance of Algorithm~\ref{alg1} for the fairness BER  problem \eqref{P}. The initial setup for the fairness BER problem simulations are the same as those used for evaluating the weighted sum BER approach discussed above.  Fig.~\ref{maxminfig1}  shows the penalty  violation $\xi$ in \eqref{constraintviolation} and the convergence behavior of  Algorithm~\ref{alg1} under  different periods $T$. It can be seen from Fig.~\ref{maxminfig1}(a) that $\xi$  converges very fast with the value decreasing to $10^{-10}$ after $34$ iterations for $T=20~\rm s$. Even when $T=40~\rm s$, the constraint is eventually satisfied within the predefined accuracy (i.e., $10^{-10}$) by $34$ iterations, which indicates that the proposed penalty-based algorithm can effectively tackle the binary scheduling constraints. In addition, in  Fig.~\ref{maxminfig1}(b), we plot the  fairness utility value versus the number of outer layer iterations. We see that the  fairness utility value increases quickly with the number of outer layer iterations for both the $T=20~\rm s$ and $T=40~\rm s$ cases, and convergence to  a fraction of the final value is achieved only 4 iterations.
\begin{figure}[!t]
	\centering
	\begin{minipage}[t]{0.45\textwidth}
		\centering
		\includegraphics[width=3.2in]{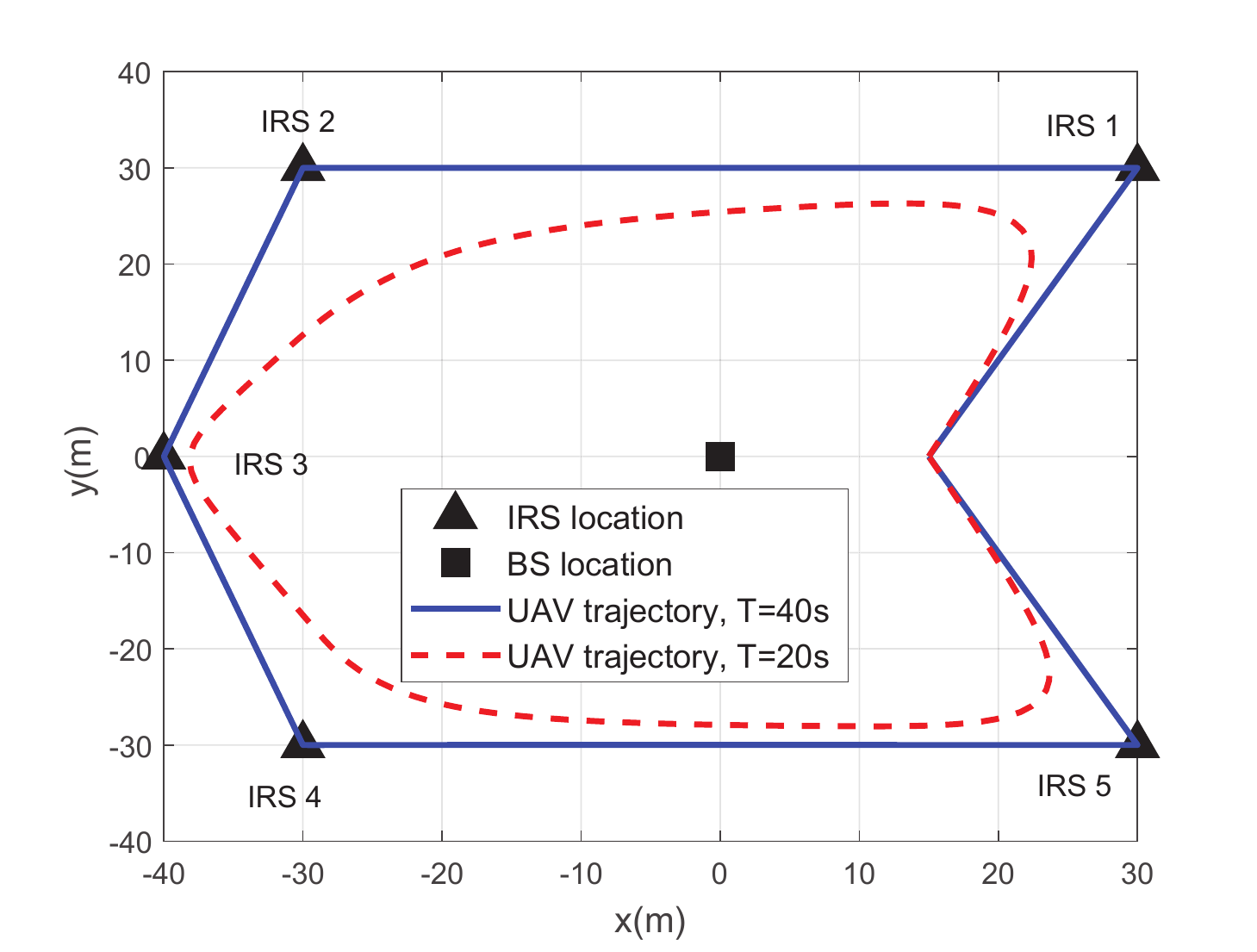}
		\caption{Optimized UAV trajectories for different  $T$.}\label{maxminfig2}
	\end{minipage}
	\begin{minipage}[t]{0.45\textwidth}
		\centering
		\includegraphics[width=3.2in]{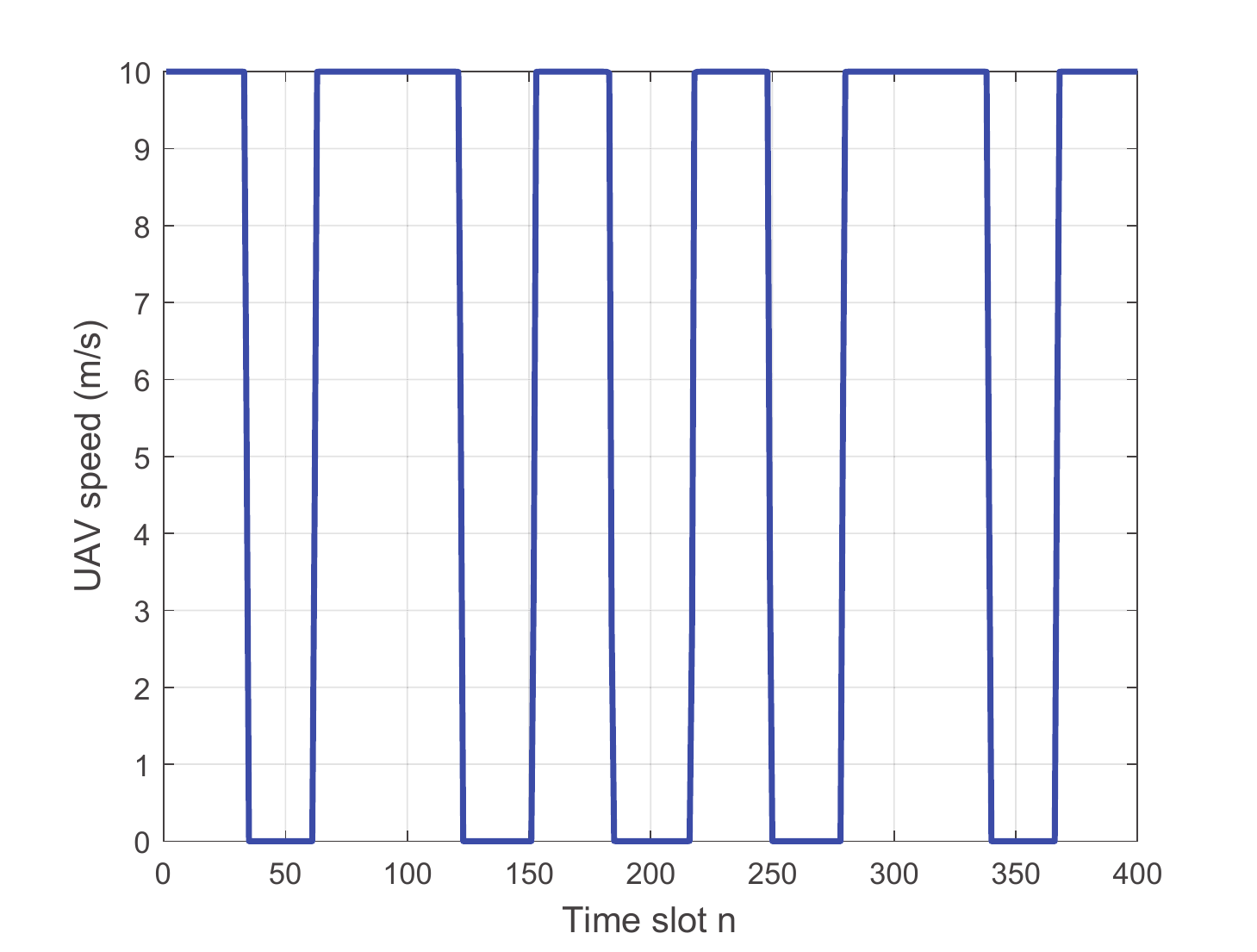}
		\caption{Optimized  UAV speed for $T=40~\rm s$.} \label{maxminfig3}
	\end{minipage}
\end{figure}

\begin{figure}[!t]
	\centering
	\begin{minipage}[t]{0.45\textwidth}
		\centering
		\includegraphics[width=3.2in]{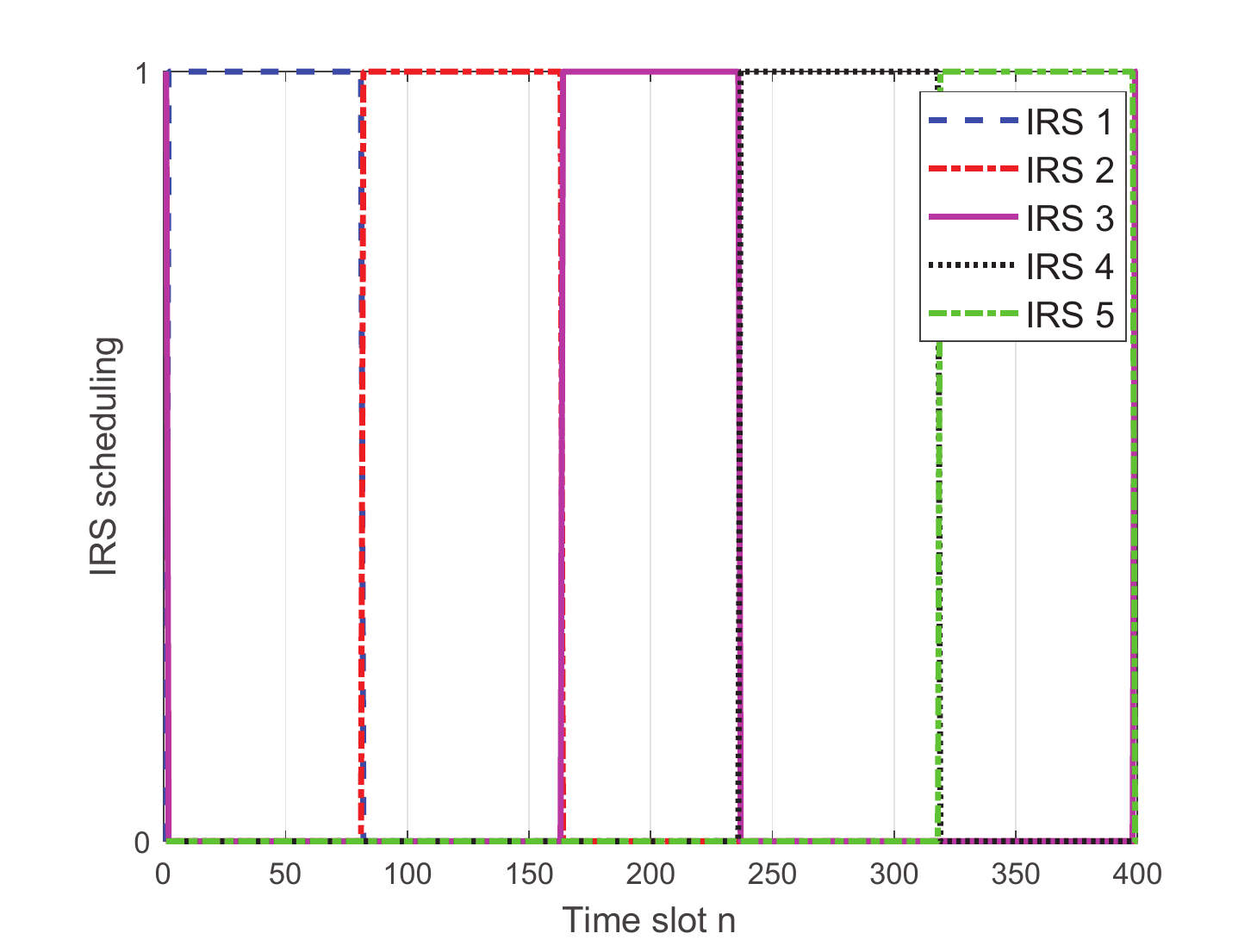}
		\caption{Optimized IRS scheduling for $T=40s$.}\label{maxminfig4}
	\end{minipage}
	\begin{minipage}[t]{0.45\textwidth}
		\centering
		\includegraphics[width=3.2in]{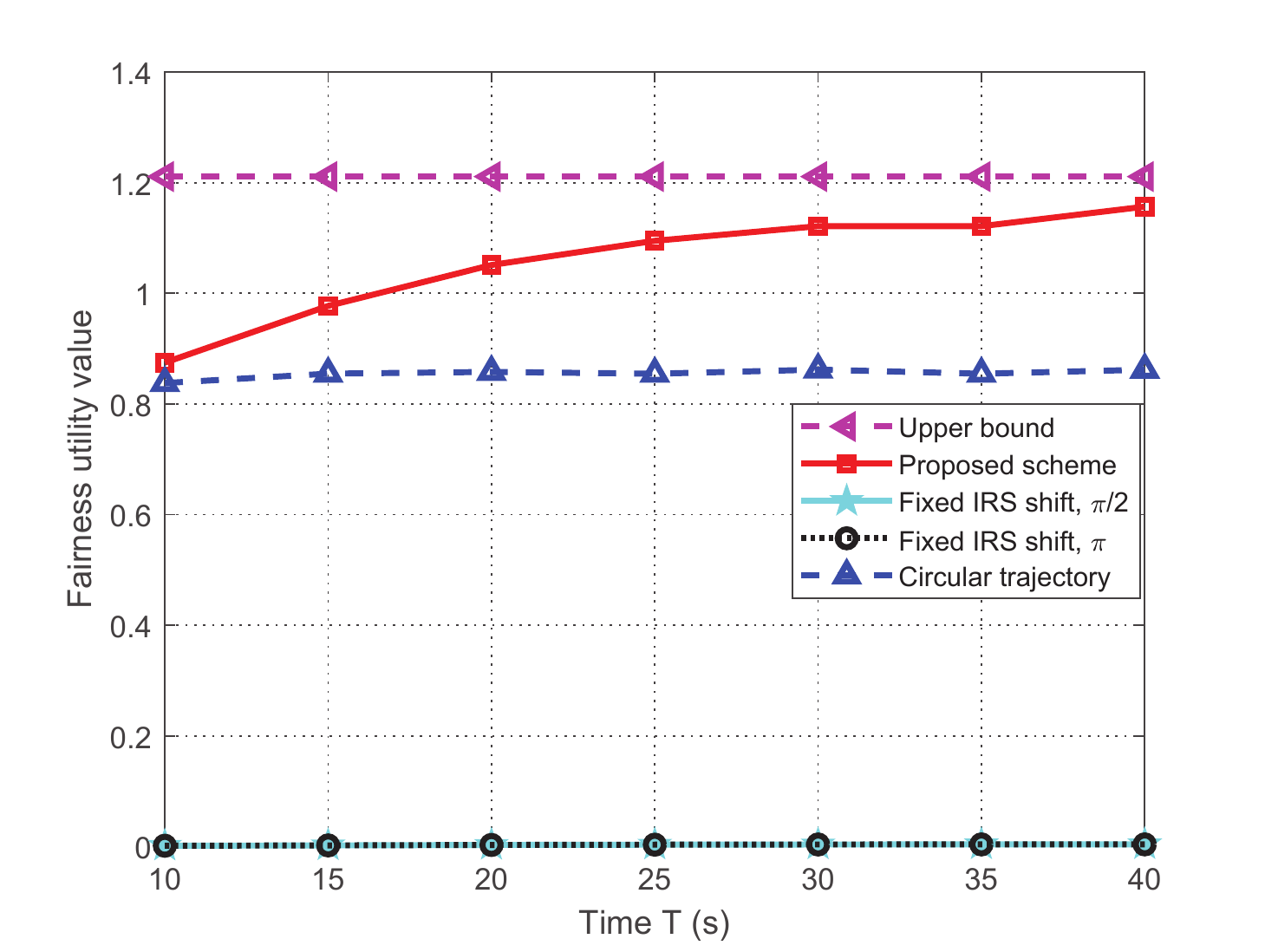}
		\caption{Average fairness utility value versus period $T$.} \label{maxminfig5}
	\end{minipage}
\end{figure}

In Fig.~\ref{maxminfig2},  the optimized trajectories obtained by  Algorithm~\ref{alg1} for $T=20~\rm s$ and $T=40~\rm s$ are plotted. As $T$ increases, the UAV exploits its mobility to adaptively enlarge and adjust its trajectory to move closer to each IRS. When $T$ becomes sufficiently large, i.e., $T=40~\rm s$, the UAV is able to sequentially visit all the IRS and stay stationary above each of them for a certain amount of time. This is expected since when the distance between the UAV and IRS is small, the length of the double channel fading propagation, i.e., the UAV-IRS-BS link, will be reduced, thus improving the IRS transmission SNR. To see this more clearly, Fig.~\ref{maxminfig3} plots the UAV speed for the case when $T=40~\rm s$. We see that the UAV flies either with maximum or zero speed, indicating that the UAV flies with maximum UAV speed to move closer to the IRS, and then remains stationary above it as soon as possible. Additionally, we observe in Fig.~\ref{maxminfig4} that the IRS sequentially communicates with each UAV  to experience better channel conditions, and the scheduling results are indeed binary, which demonstrates that the constraints in \eqref{SectionIII_7} and \eqref{SectionIII_8} are satisfied by the proposed Algorithm~\ref{alg1}.

In Fig.~\ref{maxminfig5}, we study the  average fairness utility value versus period $T$ for our proposed scheme compared with the same benchmarks as those considered for the weighted sum BER problem in Fig.~\ref{WSRfig5}.  Fig.~\ref{maxminfig5} shows that the   fairness utility value for the circular trajectory is constant regardless of the period  $T$ due to the time-invariant air-to-ground channels. In contrast, the   fairness utility value  achieved by the proposed scheme increases with $T$, which further demonstrates  the benefits of leveraging the UAV mobility.
% It is first observed from Fig.~\ref{maxminfig5} that our proposed algorithm substantially outperforms the other methods in terms of max-min rate. This is expected since an optimized UAV trajectory can establish better channel conditions for the IRS, which significantly increases IRS transmission rate. In addition, by adjusting the IRS phase shifts to align the cascaded AoA and AoD with the UAV-BS link, i.e., as shown in Theorem~2,  the SNR of the UAV-IRS-BS link will be significantly increased. 
The calculation of the upper bound for the  fairness BER  problem is different from that for the weighted sum BER problem.
When $T$ is sufficiently large, it can be assumed that the amount of time each IRS served is equal.  As for the case when the  UAV hovers above the IRS, an upper bound for the fairness BER problem  can be obtained by solving the following problem 
\begin{subequations} \label{simulationupperbound}
\begin{align}
&\mathop {\max }\limits_{{x_k} \ge 0,{R^{{\rm{upper}}}}} {R^{{\rm{upper}}}}\\
&{\rm{s}}{\rm{.t}}{\rm{.}}{\kern 1pt} {\kern 1pt} {x_k}{\log _2}\left( {1 + \frac{{LP\left( {{c_{k,1}} + {c_{k,3}}} \right){\beta _0}}}{{{\sigma ^2}{{\left( {{H_u} - {H_s}} \right)}^{{\alpha _1}}}}}} \right) \ge {R^{{\rm{upper}}}},\forall k,\\
&\qquad \sum\limits_{k = 1}^K {{x_k}}  = 1,
\end{align}
\end{subequations}
where the term ${\log _2}\left( {1 + \frac{{LP\left( {{c_{k,1}} + {c_{k,3}}} \right){\beta _0}}}{{{\sigma ^2}{{\left( {{H_u} - {H_s}} \right)}^{{\alpha _1}}}}}} \right)$  represents the achievable rate for the IRS when the UAV is directly above IRS $k$, and $x_k$ denotes the  travel time ratio for IRS $k$. Problem \eqref{simulationupperbound} is a linear optimization problem, and thus can be easily solved by the interior point method. 
\begin{figure}[!t]
	\centerline{\includegraphics[width=3.5in]{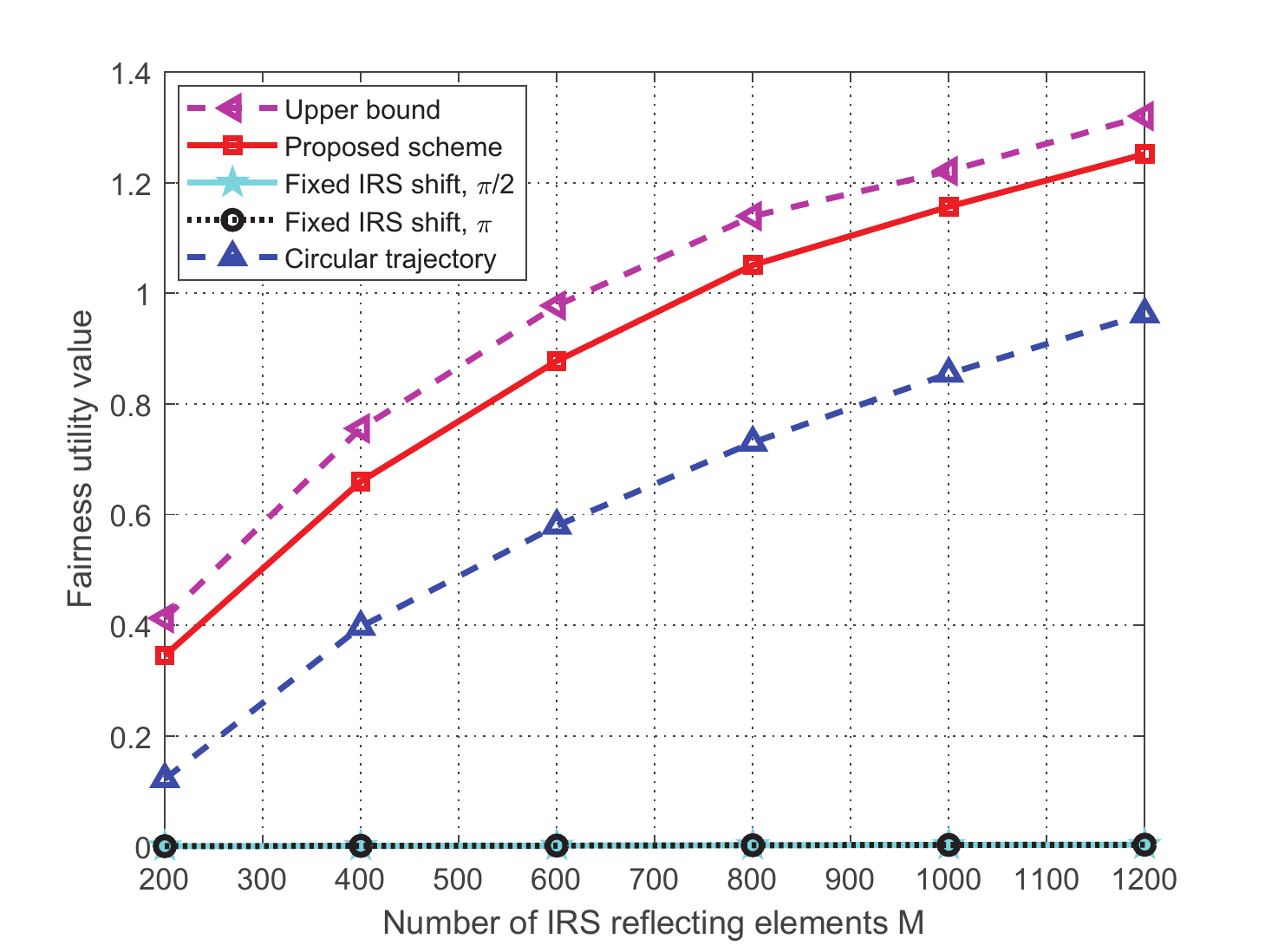}}
	\caption{Average fairness utility value versus  the number of IRS reflecting elements.} \label{maxminfig6}
\end{figure}

In Fig.~\ref{maxminfig6}, the average  fairness utility value versus the number of IRS reflecting elements is studied. We see that the performance gain of  the proposed scheme  increases as the number of IRS reflecting elements increases, since  more  reflecting elements help achieve  higher passive beamforming gain. In addition, the performance of the fixed IRS phase shift scheme is very poor, and the average fairness utility value nearly approaches zero due to the unaligned angles of the UAB-IRS-BS and UAV-BS links, which implies that the IRS phase shift must be carefully tuned.

\section{Conclusion}
In this paper, we studied a UAV-assisted IRS symbiotic radio system.
We exploited the UAV mobility to maximize the data information transferred from several IRSs to a given BS. We first studied a weighted sum BER minimization problem  by jointly optimizing the UAV trajectory, IRS phase shifts, and IRS scheduling, and proposed a low-complexity relaxation-based method to solve it. We proved that the solution to the relaxed problem provides binary scheduling results, and hence no additional operation is needed to enforce this constraint. We then considered fairness among the IRSs, and developed a fairness BER optimization problem. To handle the resulting mixed integer non-convex problem, we transformed the binary constraints into an equivalent set of equality constraints, and  proposed a penalty-based method to address the constraints. The effectiveness of this approach was justified by the numerical simulations.  Simulation results demonstrated that the system performance  can be   significantly improved by optimizing the UAV trajectory as well as the IRS phase shifts.  

\appendices
\section{Proof of Theorem~1 } \label{appendix1}
To show Theorem 1, we first define the function $f\left( z \right) = {\log _2}\left( {1 + z} \right),z \ge 0$. It can be readily checked that $f(z)$ is concave with respect to $z$. Thus, based on  Jensen's  inequality \cite{boyd2004convex}, we have ${\mathbb E}\left\{ {f\left( z \right)} \right\} \le {\log _2}\left( {1 + {\mathbb E }\left\{ z \right\}} \right)$. Therefore, the following inequality holds
\begin{align}
{\mathbb E}\left\{ {{{\bar R}_{u,k}}\left[ {n} \right]} \right\} \le& \rho {\log _2}\left( {1 + \frac{{P{\mathbb E}\left\{ {{{\left| {{h_3}\left[ n \right] + {\bf h}_{2,k}^H\left[ n \right]{{\bf\Theta} _k}\left[ n \right]{{\bf h}_{1,k}}\left[ n \right]} \right|}^2}} \right\}}}{{{\sigma ^2}}}} \right) + \notag\\
&\left( {1 - \rho } \right){\log _2}\left( {1 + \frac{{P{\mathbb E}\left\{ {{{\left| {{h_3}\left[ {n} \right]} \right|}^2}} \right\}}}{{{\sigma ^2}}}} \right).\label{appendixconst1}
\end{align}
Since the small-scale fading channel coefficients $h_3^{{\rm{NLoS}}}\left[ n \right]$, ${\bf{h}}_{1,k}^{{\rm{NLoS}}}\left[ n \right]$, and ${\bf{h}}_{2,k}^{{\rm{NLoS}}}\left[ n \right]$ are independent of each other, we  can obtain
\begin{align}
&{\mathbb E}\left\{ {{{\left| {{h_3}\left[ n \right] + {\bf{h}}_{2,k}^H\left[ n \right]{{\bf{\Theta }}_k}\left[ n \right]{{\bf{h}}_{1,k}}\left[ n \right]} \right|}^2}} \right\} = \notag\\
&\qquad{\left| {{x_{0,k}}\left[ n \right]} \right|^2} + {\mathbb E}\left\{ {{{\left| {{x_{1,k}}\left[ n \right]} \right|}^2}} \right\} + {\mathbb E}\left\{ {{{\left| {{x_{2,k}}\left[ n \right]} \right|}^2}} \right\} + {\mathbb E}\left\{ {{{\left| {{x_{3,k}}\left[ n \right]} \right|}^2}} \right\} + {\mathbb E}\left\{ {{{\left| {{x_{4,k}}\left[ n \right]} \right|}^2}} \right\}, \label{appendixconst2}
\end{align}
where ${x_{0,k}}\left[ n \right] = \sqrt {\frac{{{K_3}{\beta _3}\left[ n \right]}}{{{K_3} + 1}}} h_3^{{\rm{LoS}}}\left[ n \right] + \sqrt {\frac{{{K_1}{K_2}{\beta _{1,k}}\left[ n \right]{\beta _{2,k}}}}{{\left( {{K_1} + 1} \right)\left( {{K_2} + 1} \right)}}} {\left( {{\bf{h}}_{2,k}^{{\rm{LoS}}}\left[ n \right]} \right)^H}{{\bf{\Phi }}_k}\left[ n \right]{\bf{h}}_{1,k}^{{\rm{LoS}}}\left[ n \right]$, 

\noindent${x_{1,k}}\left[ n \right]{\rm{ = }}\sqrt {\frac{{{\beta _3}\left[ n \right]}}{{{K_3} + 1}}} h_3^{{\rm{NLoS}}}\left[ n \right]$, ${x_{2,k}}\left[ n \right]{\rm{ = }}\sqrt {\frac{{{K_1}{\beta _{1,k}}\left[ n \right]{\beta _{2,k}}}}{{\left( {{K_1} + 1} \right)\left( {{K_2} + 1} \right)}}} {\left( {{\bf{h}}_{2,k}^{{\rm{NLoS}}}\left[ n \right]} \right)^H}{{\bf{\Phi }}_k}\left[ n \right]{\bf{h}}_{1,k}^{{\rm{LoS}}}\left[ n \right]$, 

\noindent${x_{3,k}}\left[ n \right]{\rm{ = }}\sqrt {\frac{{{K_2}{\beta _{1,k}}\left[ n \right]{\beta _{2,k}}}}{{\left( {{K_1} + 1} \right)\left( {{K_2} + 1} \right)}}} {\left( {{\bf{h}}_{2,k}^{{\rm{LoS}}}\left[ n \right]} \right)^H}{{\bf{\Phi }}_k}\left[ n \right]{\bf{h}}_{1,k}^{{\rm{NLoS}}}\left[ n \right]$, and 

\noindent${x_{4,k}}\left[ n \right]{\rm{ = }}\sqrt {\frac{{{\beta _{1,k}}\left[ n \right]{\beta _{2,k}}}}{{\left( {{K_1} + 1} \right)\left( {{K_2} + 1} \right)}}} {\left( {{\bf{h}}_{2,k}^{{\rm{NLoS}}}\left[ n \right]} \right)^H}{{\bf{\Phi }}_k}\left[ {n,l} \right]{\bf{h}}_{1,k}^{{\rm{NLoS}}}\left[ n \right]$. We first calculate 
\begin{align}
{\mathbb E}\left\{ {{{\left| {{x_{2,k}}\left[ n \right]} \right|}^2}} \right\}&{\rm{ = }}\frac{{{K_1}{\beta _{1,k}}\left[ n \right]{\beta _{2,k}}}}{{\left( {{K_1} + 1} \right)\left( {{K_2} + 1} \right)}}{\left( {{\bf{h}}_{1,k}^{{\rm{LoS}}}\left[ n \right]} \right)^H}{\left( {{{\bf{\Phi }}_k}\left[ n \right]} \right)^H}{\mathbb E}\left\{ {{\bf{h}}_{2,k}^{{\rm{NLoS}}}\left[ n \right]{{\left( {{\bf{h}}_{2,k}^{{\rm{NLoS}}}\left[ n \right]} \right)}^H}} \right\}{{\bf{\Phi }}_k}\left[ n \right]{\bf{h}}_{1,k}^{{\rm{LoS}}}\left[ n \right]\notag\\
& \overset{(a)}{=} \frac{{{K_1}M{\beta _{1,k}}\left[ n \right]{\beta _{2,k}}}}{{\left( {{K_1} + 1} \right)\left( {{K_2} + 1} \right)}}, \label{appendixconst3}
\end{align}
where $(a)$ holds since ${\mathbb E}\left\{ {{\bf{h}}_{2,k}^{{\rm{NLoS}}}\left[ n \right]{{\left( {{\bf{h}}_{2,k}^{{\rm{NLoS}}}\left[ n \right]} \right)}^H}} \right\} = {{\bf{I}}_M}$, ${\left( {{{\bf{\Phi }}_k}\left[ n \right]} \right)^H}{{\bf{\Phi }}_k}\left[ n \right] = {{\bf{I}}_M}$, and 

\noindent${\left( {{\bf{h}}_{1,k}^{{\rm{LoS}}}\left[ n \right]} \right)^H}{\bf{h}}_{1,k}^{{\rm{LoS}}}\left[ n \right] = M$.
We can obtain the remaining terms as follows:
\begin{align}
{\mathbb E}\left\{ {{{\left| {{x_{1,k}}\left[ n \right]} \right|}^2}} \right\}{\rm{ = }}\frac{{{\beta _3}\left[ n \right]}}{{{K_3} + 1}}, {\mathbb E}\left\{ {{{\left| {{x_{3,k}}\left[ n \right]} \right|}^2}} \right\}{\rm{ = }}\frac{{{K_2}M{\beta _{1,k}}\left[ n \right]{\beta _{2,k}}}}{{\left( {{K_1} + 1} \right)\left( {{K_2} + 1} \right)}}, {\mathbb E}\left\{ {{{\left| {{x_{4,k}}\left[ n \right]} \right|}^2}} \right\} = \frac{{M{\beta _{1,k}}\left[ n \right]{\beta _{2,k}}}}{{\left( {{K_1} + 1} \right)\left( {{K_2} + 1} \right)}}. \label{appendixconst4}
\end{align}
In addition, we have ${\mathbb E}\left\{ {{{\left| {{h_3}\left[ {n} \right]} \right|}^2}} \right\} = {\beta _3}\left[ n \right]$. Combining all of the above results, we can directly arrive at \eqref{sectionIIprimaryrateNEW}.

\section{Proof of Theorem~2 } \label{appendix2}
Here we derive a closed-form solution for the IRS phase shifts that maximize the primary rate expression $ \hat R_{u,k}[n]$ in \eqref{sectionIIprimaryrateNEW}. We have the following inequality
\begin{small}
\begin{align}
&\left| {{x_{0,k}}\left[ n \right]} \right| = \left| {\sqrt {\frac{{{K_3}{\beta _3}\left[ n \right]}}{{{K_3} + 1}}} h_3^{{\rm{LoS}}}\left[ n \right] + \sqrt {\frac{{{K_1}{K_2}{\beta _{1,k}}\left[ n \right]{\beta _{2,k}}}}{{\left( {{K_1} + 1} \right)\left( {{K_2} + 1} \right)}}} {{\left( {{\bf{h}}_{2,k}^{{\rm{LoS}}}\left[ n \right]} \right)}^H}{{\bf{\Phi }}_k}\left[ n \right]{\bf{h}}_{1,k}^{{\rm{LoS}}}\left[ n \right]} \right|\notag\\
& = \left| {\sqrt {\frac{{{K_3}{\beta _3}\left[ n \right]}}{{{K_3} + 1}}} \exp \left( { - j\frac{{2\pi {d_3}\left[ n \right]}}{\lambda }} \right) + \sqrt {\frac{{{K_1}{K_2}{\beta _{1,k}}\left[ n \right]{\beta _{2,k}}}}{{\left( {{K_1} + 1} \right)\left( {{K_2} + 1} \right)}}} \exp \left( { - j\frac{{2\pi \left( {{d_{1,k}}\left[ n \right] - {d_{2,k}}} \right)}}{\lambda }} \right)} \right.\notag\\
&\quad\left. {\sum\limits_{m = 1}^M {\exp \left( {j\left( {\frac{{2\pi d\left( {\cos {\phi _{2,k}} - \cos {\phi _{1,k}}\left[ n \right]} \right)\left( {m - 1} \right)}}{\lambda } + {\theta _{k,m}}\left[ n \right]} \right)} \right)} } \right|\notag\\
& \mathop  \le \limits^{(a)} \left| {\sqrt {\frac{{{K_3}{\beta _3}\left[ n \right]}}{{{K_3} + 1}}} \exp \left( { - j\frac{{2\pi {d_3}\left[ n \right]}}{\lambda }} \right)} \right| + \left| {\sqrt {\frac{{{K_1}{K_2}{\beta _{1,k}}\left[ n \right]{\beta _{2,k}}}}{{\left( {{K_1} + 1} \right)\left( {{K_2} + 1} \right)}}} \exp \left( { - j\frac{{2\pi \left( {{d_{1,k}}\left[ n \right] - {d_{2,k}}} \right)}}{\lambda }} \right)} \right.\notag\\
&\left. { \times \sum\limits_{m = 1}^M {\exp \left( {j\left( {\frac{{2\pi d\left( {\cos {\phi _{2,k}} - \cos {\phi _{1,k}}\left[ n \right]} \right)\left( {m - 1} \right)}}{\lambda } + {\theta _{k,m}}\left[ n \right]} \right)} \right)} } \right|,
\end{align}
\end{small}
where $(a)$ is due to the triangle inequality, which holds with equality if and only if $ - j\frac{{2\pi {d_3}\left[ n \right]}}{\lambda } =  - j\frac{{2\pi \left( {{d_{1,k}}\left[ n \right] - {d_{2,k}}} \right)}}{\lambda } + j\frac{{2\pi d\left( {\cos {\phi _{2,k}} - \cos {\phi _{1,k}}\left[ n \right]} \right)\left( {m - 1} \right)}}{\lambda } + {\theta _{k,m}}\left[ n \right]$, $\forall m$. This indicates  that the $m$th phase shift at IRS $k$ should be tuned such that the phase of the signal that passes through the UAV-IRS and IRS-BS links is aligned with that of the signal over the UAV-BS direct link to achieve coherent
signal combining at the BS. Thus, we can obtain the closed-form IRS phase shift expression in \eqref{theorem2}. In addition, it can be easily checked that ${\theta^{\rm opt} _{k,m}}\left[ n \right]$ in \eqref{theorem2} is also the optimal solution that maximizes the IRS reflecting rate in~\eqref{sectionIIIRSrateNEW}. This  completes the proof of Theorem 2.

\section{Proof of Theorem~3 } \label{appendix3}
It can be readily  verified that problem \eqref{wsrP1-1}  satisfies Slater's condition, and thus strong duality holds and its optimal solution can be obtained by solving its dual problem \cite{boyd2004convex}. Specifically, we first introduce the dual variables $\{\lambda[n]\ge 0\}$ associated with the primary rate
constraints \eqref{wsrPconst1NEW}, and derive the partial Lagrangian of problem \eqref{wsrP1-1} as follows
\begin{align}
{\cal L}\left( {{a_k}\left[ n \right],\lambda \left[ n \right]} \right)&= \sum\limits_{k = 1}^K {{w_k}\sum\limits_{n = 1}^N {{a_k}\left[ n \right]{\cal F}\left( {{\gamma _k}\left[ n \right]} \right)} }  + \sum\limits_{n = 1}^N {\lambda \left[ n \right]\left( {\sum\limits_{k = 1}^K {{a_k}\left[ n \right]{R_{u,k}}\left[ n \right] - {R_{{\rm{th}}}}} } \right)} \notag\\
&= \sum\limits_{n = 1}^N {\left( {\sum\limits_{k = 1}^K {\left( {{w_k}{\cal F}\left( {{\gamma _k}\left[ n \right]} \right) + \lambda \left[ n \right]{R_{u,k}}\left[ n \right]} \right){a_k}\left[ n \right] - \lambda \left[ n \right]{R_{{\rm{th}}}}} } \right)}.  \label{wsrP1-1Lagrangian}
\end{align}
The  Lagrange dual function of \eqref{wsrP1-1} is  defined as
\begin{subequations} \label{wsrP1-1dualfunction}
	\begin{align}
	&g\left( {\lambda \left[ n \right]} \right) = \mathop {\max }\limits_{{a_k}\left[ n \right]} {\cal L}\left( {{a_k}\left[ n \right],\lambda \left[ n \right]} \right) \\
	&\qquad \qquad{\rm s.t.}~\eqref{Pconst2},\eqref{Pconst3NEW}.
	\end{align}
\end{subequations}
It can be seen that the dual function \eqref{wsrP1-1dualfunction} can be divided into $N$ subproblems that can be solved in parallel.  The $n'$-th subproblem of  \eqref{wsrP1-1dualfunction} can be written as
\begin{subequations} \label{wsrP1-1dualfunctionsubproblem}
\begin{align}
&\mathop {\max }\limits_{{a_k}\left[ {n'} \right]} \sum\limits_{k = 1}^K {\left( {{w_k}{\cal F}\left( {{\gamma _k}\left[ n' \right]} \right) + \lambda \left[ {n'} \right]{R_{u,k}}\left[ {n'} \right]} \right){a_k}\left[ {n'} \right] - \lambda \left[ {n'} \right]{R_{{\rm{th}}}}} \\
&{\rm{s}}{\rm{.t}}{\rm{.}}{\kern 1pt} {\kern 1pt} {\kern 1pt} 0 \le {a_k}\left[ {n'} \right] \le 1,\forall k,\\
&\qquad\sum\limits_{k = 1}^K {{a_k}\left[ {n'} \right]}  \le 1.
\end{align}
\end{subequations}
%Obviously, for any given $\lambda[n']\ge0$, we have ${{w_k}f\left( {{\gamma _k}\left[ n' \right]} \right) + \lambda \left[ {n'} \right]{R_{u,k}}\left[ {n'} \right]}>0$. 
It can be easily derived that the optimal solution $a^{\rm opt}_k[n']$ that maximizes \eqref{wsrP1-1dualfunctionsubproblem} is either $a^{\rm opt}_{k'}[n']=1$ or $a^{\rm opt}_{k}[n']=0$ for  $k \ne k$, where subscript $k'$ corresponds to the index that maximizes ${{w_k}{\cal F}\left( {{\gamma _k}\left[ n' \right]} \right) + \lambda \left[ {n'} \right]{R_{u,k}}\left[ {n'} \right]}$ among all $k \in \left\{ {1, \ldots ,K} \right\}$. This also holds for the case that there are  more than two IRS that have  the same  maximum value of ${{w_k}{\cal F}\left( {{\gamma _k}\left[ n' \right]} \right) + \lambda \left[ {n'} \right]{R_{u,k}}\left[ {n'} \right]}$ among all $k \in \left\{ {1, \ldots ,K} \right\}$. This thus completes the proof of Theorem 3.

\bibliographystyle{IEEEtran}
\bibliography{IRSUAV}
\end{document}